\documentstyle[epsf,axodraw]{aipproc}
\def\upr#1{\uppercase\expandafter{#1}}
\begin{document}
\author{Fizuli Mamedov}
\address{Department of Physics, State University of New York, 
Buffalo, New York 14260} 
\title{Weak Boson Production Amplitude Zeros; Equalities of the 
Helicity Amplitudes} 
\maketitle
\begin{abstract}
{We investigate the radiation amplitude zeros exhibited by many Standard 
Model amplitudes for triple weak gauge boson production processes. We show that $WZ\gamma$ production amplitudes have especially rich structure in terms of zeros, these amplitudes have zeros originating from several different sources. It is also shown that 
TYPE I current null zone is the special case of the equality of the specific helicity amplitudes.}
\end{abstract}

\begin{section}{Introduction}
\indent The SM amplitudes for processes with the neutral gauge boson(s) emission exhibit zeros. Either the distribution of the scattering angle 
 contains zeros, or the helicity amplitudes completely vanish for 
 certain polarization (or momentum) combinations of the  particles which
 participate in the process. In this work we review the theoretical basis of these radiation amplitude zeros, and discuss the experimental aspects of this phenomenon in weak boson production processes.

In the works related  to the analysis of the unified theories of the electromagnetic and weak interactions it was pointed out that there was a connection
between  the values of the magnetic moment of the vector-bosons and 
possible structures of these theories \cite{W_moment_1}, \cite{W_moment_2}.
In the early days of the SM, the processes
$ pp~(p\bar p) \to W^{\pm}+\gamma+X~$
were suggested for the measurement of the magnetic moment of the $W$ boson, where the radiation amplitude zero was encountered first \cite{Rad_0}.  We discuss this zero in Section \upr{\romannumeral 2}. The observed zeros of these production processes belong to the so-called TYPE I zeros. We consider the conditions 
which have to be fulfilled in order that TYPE I radiation zeros can occur in Section \upr{\romannumeral 3}. 
We describe how these radiation zeros can be explained 
as a result of the factorization of 
the amplitudes as well as a consequence of the decoupling 
theorem. TYPE I zeros have two different forms, $charge~null~zone$
and $current~null~zone$ zeros. We also give a few examples for 
the recently discovered TYPE II zeros. Section \upr{\romannumeral 4} is 
devoted to the discussion of many interesting zeros, which occur in 
electroweak production processes.  First, we discuss the zeros in 
$W\gamma$ and $W\gamma \gamma$  production.
Here, we also briefly discuss the  equalities of the values of the specific helicity amplitudes, which are   responsible for 
the current null zone zeros. Next, we consider  
the $WZ$ production zeros. Some of the $WZ$ production amplitude zeros were
observed only recently and they can not be attributed to any type of the zeros discovered earlier. Special attention is paid to a discussion of the full set
of zeros in the $WZ$ production process, since they directly relate to the zeros of two other production processes, the zeros in  $q \bar q \to WZ\gamma$ and $q \bar q \to WZZ$, which are discussed subsequently. The amplitudes in the  $W^{\pm} Z 
\gamma$ case have an especially  rich structure in terms of zeros, revealing 
three different types of zeros. We also review the zeros in
 $WH\gamma$ production. The existence of the charge null zone
requires the same sign for all the nonzero charges of the particles participating in the process and therefore processes, where only  
 neutral electroweak gauge bosons are produced, can only exhibit 
the current null zone TYPE I zeros. We discuss several of these processes
in terms of the radiation zeros together with the helicity amplitudes relations  mentioned earlier in more detail at the end of this section. Finally, we present a summary table of the zeros, considered throughout Section \upr{\romannumeral 4}.  
\end{section}
\begin{section}{Radiation zeros in $W\gamma$ Production}
In this section we briefly discuss the radiation amplitude zeros phenomenon associated with the $p\bar p \to W^{\pm}+\gamma$   collision process\footnote{For brevity, we will not write `$+X$' addition, denoting backgrounds to the considered processes, hereafter.} in order to illustrate several important theoretical and experimental features of this phenomenon.
The  parton level amplitudes responsible for this process  exhibit zeros in the distribution of the scattering angle \cite{Rad_0}.  
\begin{figure}
\begin{center}
\epsfysize=3.85in
\epsffile{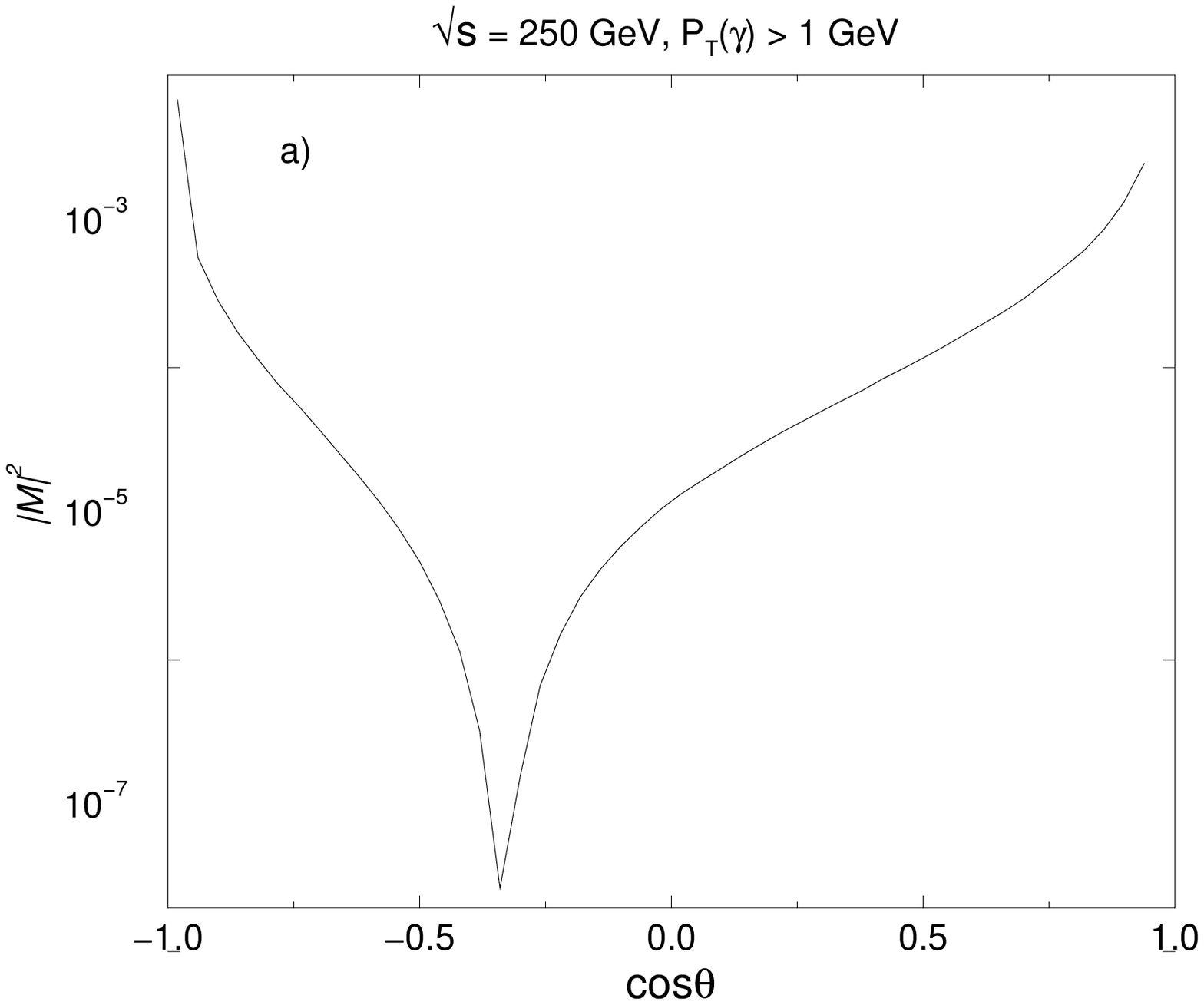}
\epsfysize=3.85in
\epsffile{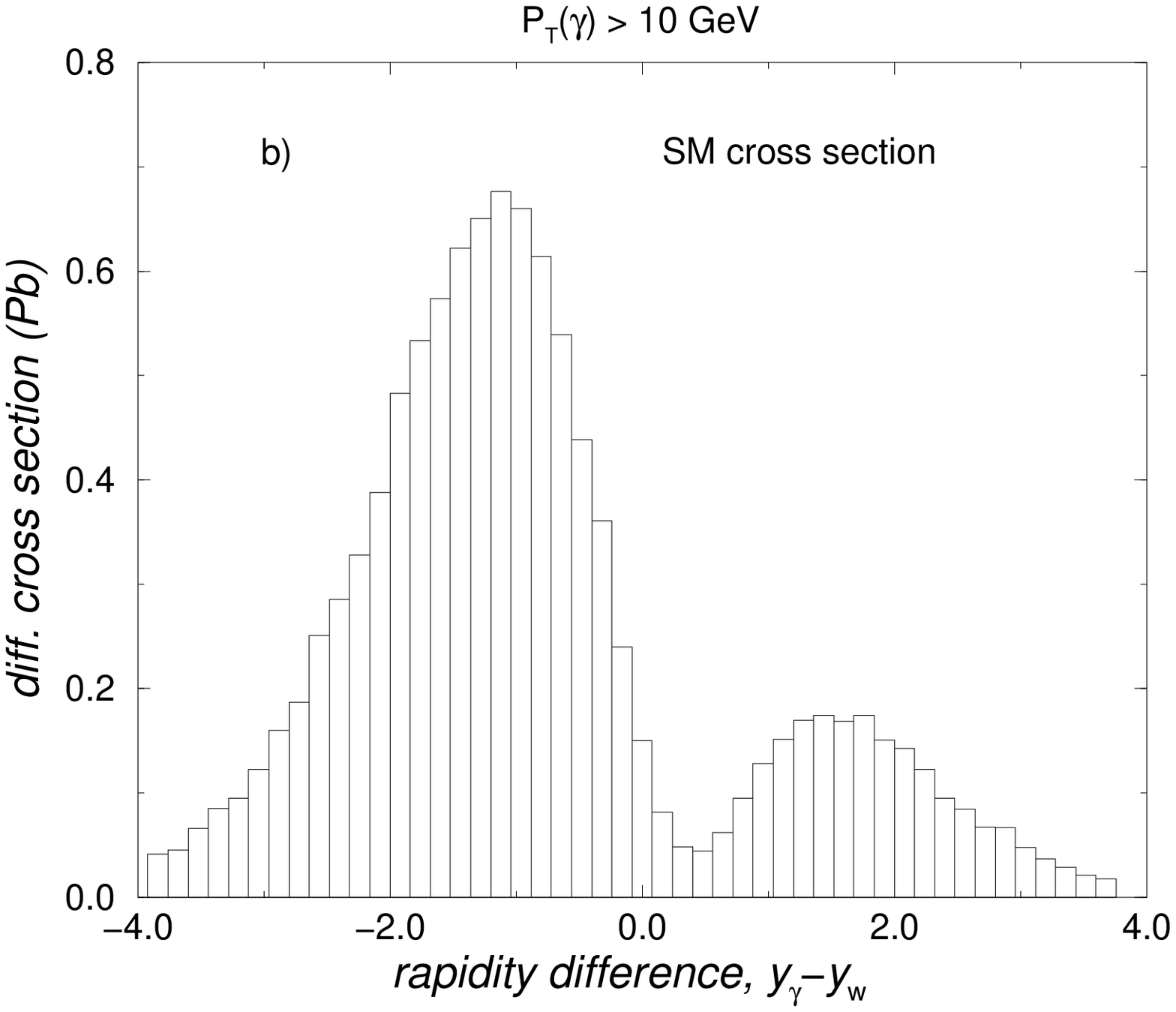} 
\end{center}
\caption{a) The zero in the angular distribution for $d \bar u \to W^-\gamma$ production for a center of mass energy of quark-antiquark system $\sqrt{s}$ = 250 GeV, 
b)  Rapidity difference distribution for the $p\bar p
\to W^{-}\gamma$ process at the Tevatron (2 TeV).}
\label{wa}
\end{figure}
The position of the zero depends only on the charge of the quark (no helicity dependence):
\begin{equation}
\label{Eq:rad0}
\cos\theta = -(1+2Q_{i})~,
\label{eq:zero}
\end{equation}
where $\theta$ is the angle between the $W^{-}$ and the $d$ quark, if one consideres  $d \bar u \to W^- \gamma$ process. $Q_i$ is the electric charge of the quark in units of the proton charge $e$.

The value of $\cos\theta=-\frac{1}{3}$ (see Fig. \ref{wa}a), obtained from 
Eq.~(\ref{Eq:rad0}), is in fact, characteristic for some other SM based
process amplitudes too, as we will discuss later.  

From the experimental point of view, the Tevatron collider 
($p\bar p$) is especially well suited to observe the radiation zero
predicted in 
$W\gamma$ production. Sea quark effects tend to wash out the dip caused
by the radiation zero. At Tevatron energies, valence quark effects
dominate and this effect is not a problem. As a result, the radiation zero 
leaves a clear signature. This is shown in Fig. \ref{wa}b where 
we display the distribution\footnote{We use histograms for 
the collider process distributions to distinguish them from the parton 
level curves and to get neat curves for the rapidity distributions at 
the relatively low Monte Carlo statistics. The histograms are also helpful to see the `bin structure' of the numerical results for the differential cross sections. In all the graphs, not using the histograms, the horizontal axes are divided into fifty equal bins.}
of the difference between the rapidities of
the $W$ boson, $y_W$, and the photon, $y_\gamma$. The dip at
$y_{\gamma}-y_{W}\approx 0.3$ is due to the radiation zero~\cite{Baur_1}.

The boost invariant quantity $y_{\gamma}-y_{W}$ contains the same
information as the $\cos\theta$ distribution. The $y_{\gamma}-y_{W}$
distribution is very similar to the distribution of the rapidity
difference between the photon and the charged lepton originating from
the $W$ decay, which can be readily observed. This is due to the $V-A$
nature of the $W$ fermion antifermion coupling and the fact that $W$'s
in $W\gamma$ production in the SM are strongly polarized: 
the dominant helicity of the $W^{\pm}$ boson in SM $W^{\pm}\gamma$ 
production is $\lambda_{W}=\pm 1$ \cite{Baur_1}.
\end{section}

\begin{section}{Radiation Amplitude Zeros and  Gauge Fields}
The amplitude zero observed in $W\gamma$ production belongs to 
the family of  `TYPE~I' zeros  which can be explained as 
a consequence of 
the factorization of the amplitude, shown soon after they were first 
discussed in the literature~\cite{Brown_1}-\cite{Brown_2}. 

The scattering amplitude for the above mentioned  process can be
obtained starting from the vertex (source graph)  which describes 
the interaction of
the charged particles, attach a photon to each charged particle 
leg in turn and add all diagrams, as schematically depicted below:
\SetScale{0.8}{
\begin{center}
\begin{picture}(500,120)(-190,-50)
\ArrowLine(-140,50)(-190,50)
\ArrowLine(-225,85)(-190,50)
\ArrowLine(-225,15)(-190,50)
\Text(-100,40)[l]{$\Rightarrow$}
\ArrowLine(-10,50)(-60,50)
\ArrowLine(-95,85)(-60,50)
\ArrowLine(-95,15)(-60,50)
\Photon(-84.75,74.75)(-59.75,74.75){4}{5}
\Vertex(-85,74.75){2}
\Text(10,40)[l]{+}
\ArrowLine(110,50)(60,50)
\ArrowLine(25,85)(60,50)
\ArrowLine(25,15)(60,50)
\Photon(35.25,24.25)(60.25,24.25){4}{5}
\Vertex(34,24.25){2}
\Text(100,40)[l]{+}
\ArrowLine(230,50)(180,50)
\ArrowLine(145,85)(180,50)
\ArrowLine(145,15)(180,50)
\Photon(215,50)(215,75){4}{5}
\Vertex(215,50){2}
\ArrowLine(-190,-20)(-140,-20)
\Text(-100,-17)[l]{$\Rightarrow$}
\Text(-85,-17)[l]{{$charged~ particle~ lines$}}
\Photon(-190,-44)(-165,-44){4}{5}
\Text(-110,-37)[l]{$\Rightarrow$}
\Text(-90,-37)[l]{{$photon~(gauge~ boson)~lines$}}
\end{picture}
\end{center}
}
One can show that  the amplitude (for particles of any 
spin) can be written in the form 
\begin{equation}
M_{\gamma} =  \sum_{i}\frac{A_{i}B_{i}}{C_{i}}~, 
\end{equation}
where $A_{i}$ and $B_{i}$ are factors which depend on the charge and
polarization. $C_i$ represent the particle propagators. 
This leads to the factorization of the amplitude into 
separately charge dependent and polarization dependent factors:
\begin{equation}
\sum_{i}\frac{A_{i}B_{i}}{C_{i}}=f( A_{i},C_{i})\cdot g( B_{i},C_{i})~.
\end{equation}
The factorization of the amplitudes holds for any 
\textit{\textbf{gauge theory based vertex}} with no restriction on 
the number of particles, due to the relation between the photon 
(gauge boson) coupling and Poincare invariance. For the complete tree 
level amplitude for a source graph $V_{G}$ consisting of a 
\textit{single vertex} (no internal lines)
\begin{equation}
M_{\gamma}(V_{G}) = \sum \frac{Q_{i}J_{i}}{p_{i}\cdot q}~,
\end{equation}
where $J_{i}$ are 'the vertex currents', arising from inserting 
the current $j_{i}$ into the $i^{th}$ leg of the vertex, with
\begin{equation}
\label{Eq:j}
j=j_{conv}+j_{spin}+j_{cont}+j_{YM}~.
\end{equation}
The convection current, $j_{conv}=p \cdot q$, corresponds to the first-order 
space-time translation of the given leg's wave function. 

The spin current is a first-order Lorentz transformation of the wave 
function. For the Lorentz transformation 
\begin{equation} 
\Lambda_{\mu\nu}=g_{\mu\nu}+\lambda \omega_{\mu\nu}~,
\end{equation}
the spinor wave function $\psi$ transforms as
\begin{equation} 
\psi^{\prime}(x^{\prime})=S(\Lambda) \psi(x)~,
\end{equation}
where 
\begin{equation} 
\omega_{\mu\nu}=q_{\mu} \epsilon_{\nu}-\epsilon_{\mu} q_{\nu}~,
\end{equation}
$\lambda$ is an infinitesimal length, $x^{\prime}=\Lambda x$, and
\begin{equation} 
S(\Lambda)=1-\frac{i}{4} \lambda \sigma_{\mu \nu} \omega^{\mu\nu}~.
\end{equation}
Therefore for a spinor
\begin{equation} 
j_{spin}=\frac{i}{4} \lambda \sigma_{\mu \nu} \omega^{\mu\nu}~.
\end{equation}

\indent The contact $j_{cont}$ and Yang-Mills  $j_{YM}$ currents result from the transformations
of the single derivative couplings and Yang-Mills vertices,
correspondingly \cite{Brown_1}.

 $J_{i}$  depend on the polarizations, but not 
on the charges of the particles, and obey the identity 
\begin{equation}
\sum J_{i}=0
\end{equation}
 as a result of Poincare and  Yang-Mills symmetries \cite{Brown_1}.

 Thus the vertex amplitude $M_{\gamma}(V_{G})$ vanishes, if
\begin{equation}
\label{Eq:chargezero}
 \frac{Q_{i}}{p_{i}\cdot q}={\rm const~,~for~all}~i~.
\end{equation}

The well-known `-$\frac{1}{3}$' zero occurring in $W\gamma$ production 
belongs to this type. The classical limit of these zeros is the vanishing 
of the dipole radiation for the system of particles with the same charge 
to mass ratio

\begin{equation}
\label{Eq:Classical0new}
Q_{i}/m_{i} = Q_{1}/m_{1}~,~{\rm for~all}~i~, 
\end{equation}
 and giromagnet coefficients, $g_i$  \cite{Brown_1}.
 
Since $\sum \delta_{i} Q_{i}=0$, the amplitude $M_{\gamma}(V_{G})$ will 
be zero, if 
\begin{equation}
\label{Eq:CurNullzone}
\delta_{i}\frac{J_{i}}{p_{i}\cdot q}={\rm const~,~for~all}~i~.
\end{equation}
These zeros correspond to the $\textit{current null zone}$. In 
the infrared limit, \mbox{Eq. (\ref{Eq:CurNullzone})} reduces to
\begin{equation}
\label{Eq:Jinfrared}
\frac{p_{i} \cdot \epsilon}{p_{i} \cdot q}={\rm const~,~for~all}~i~.
\end{equation}
Since $p_{i}\cdot q \geq 0$ for all $i$, all the convection currents, 
$p_{i} \cdot \epsilon$, have to vanish in order that 
Eq. (\ref{Eq:Jinfrared}) will be satisfied. It is easy to see that 
this implies that all the charged particles are restricted to 
the plane (a line) perpendicular to the photon polarization vector 
$~\vec \epsilon~$ for a linearly (elliptically) polarized photon, 
in the c.m. frame. This is the quantum field theory analogue of 
the classical case that there is no electric dipole radiation 
perpendicularly polarized to the scattering plane.

The null zone conditions, Eq. (\ref{Eq:chargezero}) and  
Eq. (\ref{Eq:CurNullzone}) also imply the invariance under 
the replacements
\begin{equation}
\label{Eq:Qi}
\frac{Q_i}{p_i \cdot q} \to \frac{Q_i}{p_i \cdot q}+C
\end{equation}
and 
\begin{equation}
\label{Eq:Ji}
\frac{J_i}{p_i \cdot q} \to \frac{J_i}{p_i \cdot q}+C^{\prime}~,
\end{equation}
respectively. Therefore, for a suitable choice of $C$ and $C^{\prime}$,  
the single vertex amplitude $M_{\gamma}(V_{G})$ can be written
as 
\begin{equation}
\label{Eq:M_V}
M_{\gamma}(V_{G})=\sum_{i \neq j,k} p_i \cdot q \Big(\frac{Q_i}{p_i \cdot q}-
\frac{Q_j}{p_j \cdot q}\Big)\Big(\frac{J_i}{p_i \cdot q}-
\frac{J_k}{p_k \cdot q}\Big)~,
\end{equation}
where the  null zone conditions take an explicit form.

In the case  of the source graphs with the internal lines (several source graphs),  the tree radiation amplitude $M_{\gamma}(T_{G})$  can be written as a sum over the vertices
\begin{equation}
\label{Eq:M_T}
M_{\gamma}(T_{G})=\sum_{v} M_{\gamma}[V_{G}(v)]R(v)~.
\end{equation}
 Here $R(v)$ is $T_G$ less the vertex $v$. Each of $M_{\gamma}[V_{G}(v)]$ in Eq. (\ref{Eq:M_T}) will have the same properties in terms of zeros as the single vertex amplitude $M_{\gamma}(V_{G})$, Eq. (\ref{Eq:M_V}) and therefore there exists a null zone under the condition Eq. (\ref{Eq:chargezero}) or Eq. (\ref{Eq:CurNullzone}), similar to the single vertex case. 

From the conditions above we see that the null zones connect 
intrinsic (charge, spin) and space-time properties (Poincare
transformation) of the particles involved. This makes it possible to use 
them in analysing the structure of the SM.

Radiation zeros can also be explained as a consequence of the 
decoupling theorem~\cite{Brown_4}, \cite{Brown_5}. The wave function of a
system of particles in an external Yang-Mills field can be 
written as
\begin{equation}
\Psi(x)=ULT\chi(x)~,
\end{equation} 
where $\chi(x)$ is the free solution of the field equations 
($Q=0$, no gauge boson emission), and 
$ULT$ is the product of the local gauge ($U$), Lorentz ($L$), and 
displacement ($T$) transformations. The null zone condition
\begin{equation}
\prod_{i}(ULT)_{i}=1
\end{equation}
leads to the charge null zone condition discussed above.

From the condition for the charge null zone we conclude that 
(since $p_{i} \cdot q \geq 0$)
\begin{equation}
\label{Eq:Qsign}
\frac{Q_{i}}{Q_{j}} \geq 0~, ~{\rm for~all}~i,j~.
\end{equation} 
Notice that the zeros will not necessarily be in the physical range of 
the parameters ($-1 \le \cos\theta \le 1$ in the case discussed here).

In $2\to 2$ scattering processes, where one of the final 
particles is a massless neutral gauge boson, in the relativistic 
limit, the zeros occur at the angle
\begin{equation}
\cos\theta = \frac{Q_{1}-Q_{2}}{Q_{2}+Q_{1}}~,
\end{equation}  
where $Q_{1}$ and $Q_{2}$ are the charges of the initial state particles.

For the reaction 
$d \bar u \to W^{-} \gamma$
we indeed get $\cos\theta = -\frac{1}{3}~$, consistent with the result
from a direct computation of the matrix elements.

One can also consider TYPE I zeros in supersymmetric extensions of the SM. The  emission/absorption of the gauginos will also be 
associated with radiation zeros \cite{Robinet_1}-\cite{DELANEY_1}.

The presence of amplitude zeros requires a gyro-magnetic 
factor of $g=2$. Any anomalous $WW\gamma$ coupling changes the value of 
$g$ and destroys the radiation zero \cite{Brown_1}, \cite{Brown_7}, \cite{Brown_8}.

Recently another type of the zeros (TYPE~II) was discovered 
 (\cite{Heyssler_1}-\cite{Stirling_1})
in the physical phase space range for the processes~
\begin{equation}
e^{+}u \to e^{+}u + \gamma~,\qquad e^{+}d \to e^{+}d + \gamma~,
\end{equation}
and
\begin{equation}
q \bar q \to W^{+}W^{-} \gamma~.
\end{equation}
TYPE II zeros occur only if the emitted photons are located in the 
scattering plane. The theory, revealing the underlying symmetry of the production process amplitudes responsible for this zero has yet to be offered. 
\end{section}
\begin{section}{Amplitude zeros in weak boson production processes}
In this section we discuss, somewhat systematically, the amplitude zeros 
exhibited by  electroweak production processes\footnote{For more details of the zeros, analyzed in the earlier works, see references at beginning of their discussions.}. We also show that the current null zones can be considered as a consequence of the equality of the specific helicity amplitudes. QCD corrections are expected to give noticable contributions to the distributions. However,
we do not consider them in this work, since the objective of our analysis is
to investigate the important aspects of the radiation amplitude zeros phenomenon for the weak boson production amplitudes.  MadGraph \cite{Stelzer_1} was used in the analysis of the  production amplitudes.

In our calculations we have used $M_W=80.35$ GeV, $M_Z=91.18$ GeV, $M_H=150.0$ GeV for the W boson, Z boson and Higgs masses, respectively. The values of the coupling constants were taken at the $W$ boson mass scale:
\begin{equation}
\alpha(M_W)=1/128~.
\end{equation}
\begin{subsection}{$W \gamma$, $W \gamma \gamma$ production zeros}
\indent Besides the `-$\frac{1}{3}$' zeros discussed in Section \upr{\romannumeral 2},  the $q \bar q^{\prime} \to W \gamma$ amplitudes also exhibit current null zone zeros 
\cite{Cortes_1}, \cite{Brown_3}.
The amplitudes vanish \textit{for all angles} within the SM, when the W boson
is longitudinally polarized and the photon polarization vector $~ \vec \epsilon~$
is perpendicular to the scattering plane in the c.m. frame. 
 $W \gamma$ production as well as several other weak boson production processes 
exhibit another type of the current null zone at the 90$^{\circ}$ scattering
angle. We will not discuss those zeros here. The complete description of them
can be found in \cite{Brown_3}. 

A radiation zero can occur not only when one massless neutral gauge boson
is emitted, but also if two or more are radiated, provided that the neutral
gauge bosons are all collinear \cite{Brown_1}, \cite{Brown_8}. This is 
in agreement with 
the charge null zone condition, Eq. (\ref{Eq:chargezero}), since for 
the neutral gauge boson, $Q=0$ and $p \cdot q=0$ due to its collinearity 
to the photon (the second neutral gauge boson), and therefore there is no 
explicit violation of the condition. Here $Q$ and $p$ are the charge and 
momentum of the neutral gauge boson, respectively. The zero  is located 
at the same scattering angle as in the case where only one boson is radiated.
The charge null zone zero of  $W \gamma \gamma$ production is illustrated 
in Fig. \ref{waa}a, where we show the squared amplitude as a function of 
the photon
scattering angle for $W\gamma\gamma$ production at $\sqrt{s}=300$ GeV \cite{Baur_2}. A $p_T(\gamma)> $\mbox{5 GeV} cut has been imposed to avoid the
infrared singularities associated with photon emission.
\begin{figure}
\begin{center}
\epsfysize=3.9in
\epsffile{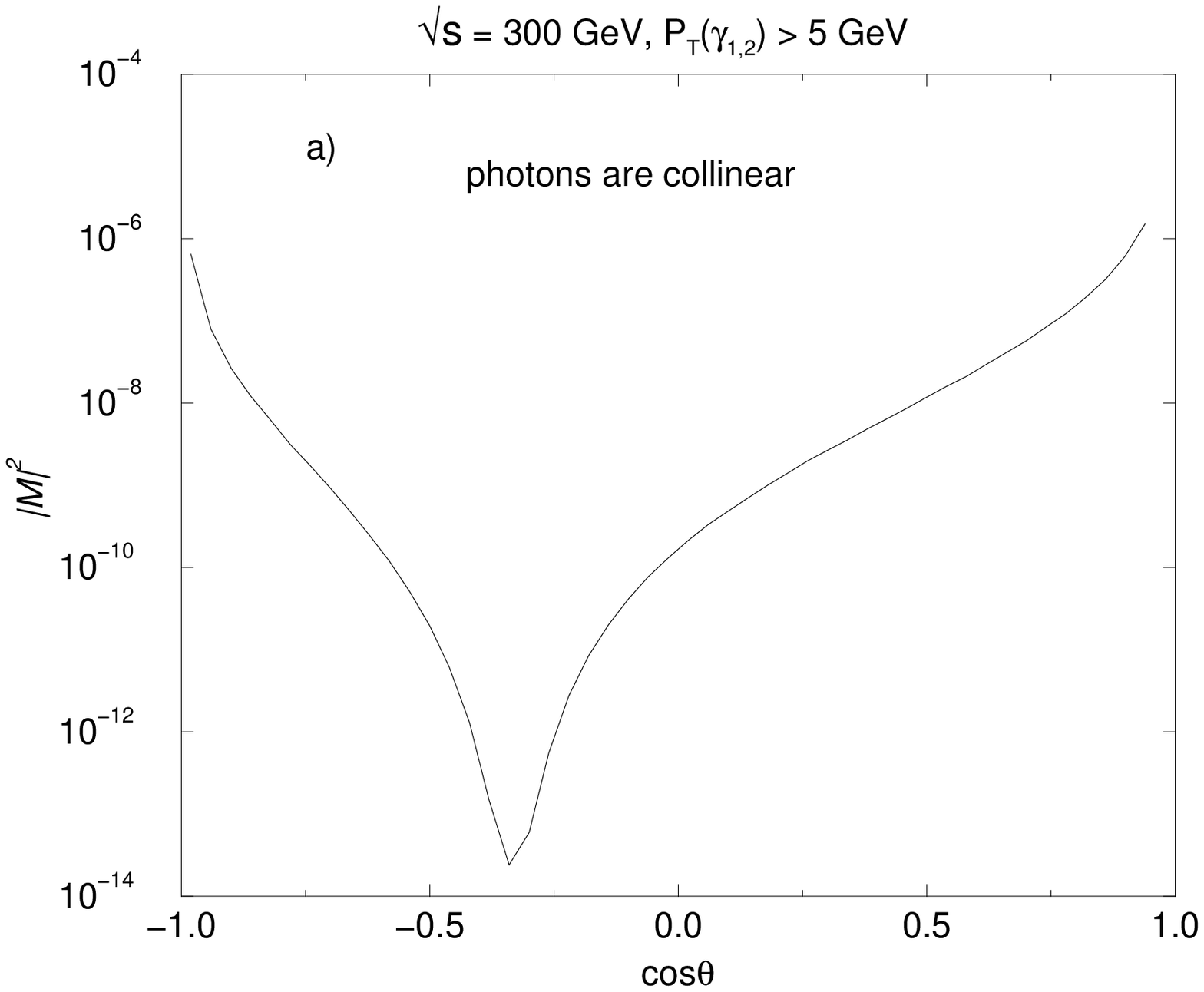}
\epsfysize=3.9in
\epsffile{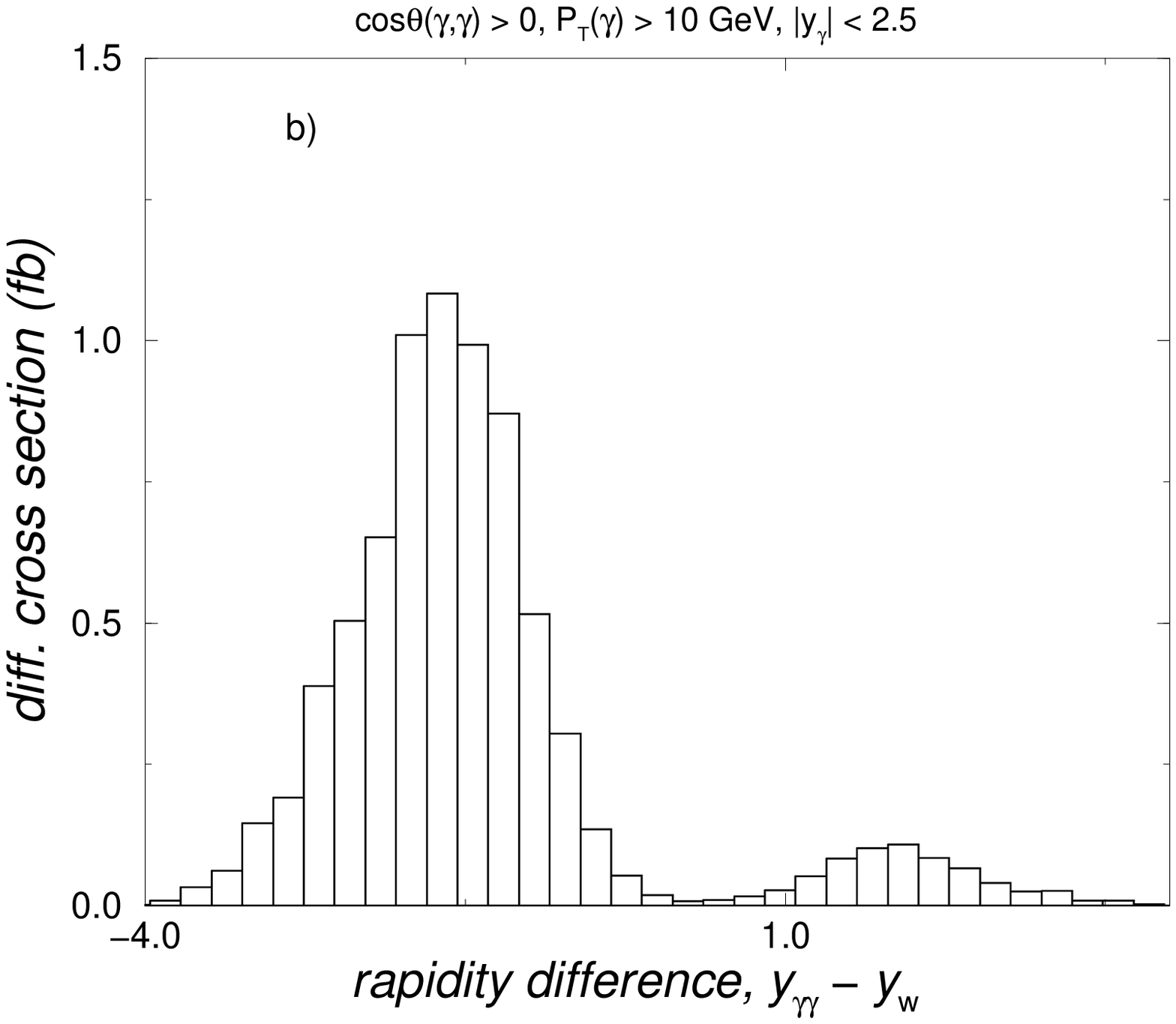} 
\end{center}
\caption{a) Radiation zero occurring in $d \bar u \to W^-\gamma\gamma$ production, b) The $~y_{\gamma \gamma}-y_{W}~$ rapidity difference distribution 
for $\cos \theta(\gamma, \gamma)>0$ in $W\gamma\gamma$ production at the Tevatron.}
\label{waa}
\end{figure}
The zero gradually vanishes for increasing values of the angle between 
the two photons. Nevertheless, there is a clear dip in the rapidity 
difference $y_{\gamma \gamma}-y_{W}$ distribution due to this zero for
\begin{equation} 
\label{Eq:PhotonAngle}
\cos \theta_{\gamma \gamma}> 0~, 
\end{equation}
as shown in Fig. \ref{waa}b, where  $\theta_{\gamma \gamma}$ is the angle between the two photons in the c.m. frame. We have imposed the following transverse
momentum  and rapidity  cuts on photon 
\cite{Han_Sobey}: 
\begin{equation} 
p_T(\gamma) > \textrm{10 GeV}~,~~~~~~~|y_{\gamma}| < 2.5~.
\end{equation}
 The helicity amplitudes for $W \gamma \gamma$ production are approximately equal in magnitude for all scattering angles and c.m. energies of the $W\gamma\gamma$ system for the following combinations of helicities
\begin{equation} 
\label{Eq:Mphoton}
M(\lambda_{W}=0,\lambda_{\gamma}=-1,\lambda_{\gamma}=-1) \approx M(\lambda_{W}=0,\lambda_{\gamma}=1,\lambda_{\gamma}=1)~,
\end{equation}
where $\lambda_{W}$ and $\lambda_{\gamma}$ are the W boson and photon
helicities. Because of the different values of both photon
helicities, these amplitudes cannot be combined into the polarization 
amplitudes, and there is no current null zone in the $W \gamma \gamma$ production case. We will 
discuss this situation in more detail for  $Z \gamma \gamma$ production 
case, where these equalities become exact.
\end{subsection}
\begin{subsection}{$WZ$ production amplitude zeros}
\indent In the case of massive neutral gauge bosons, the production 
amplitudes can still exhibit an approximate radiation zero \cite{Baur_3}, \cite{Han}. For the process
\begin{equation} 
q_{1} \bar q_{2} \to WZ
 \end{equation}
the  $M(\lambda_{W}=\pm,\lambda_{Z}= \mp)$ helicity 
amplitudes  factorize into the helicity dependent term  and the 
term, which depends on the weak boson
fermion couplings (charge dependence), but not on the helicities
of the particles (see \cite{Baur_3} for more details).
These amplitudes exhibit weakly energy dependent zeros, 
which at high energies, $\sqrt {s} \gg M_{W,Z}~$, are located at
\begin{equation} 
\label{Eq:Xtermzero}
\cos\theta = (g_{-}^{q_{1}}+g_{-}^{q_{2}})/(g_{-}^{q_{1}}-g_{-}^{q_{2}})~,
\end{equation}
where the $g_{-}^{q_{i}}$ (i=1,2) are the left-handed couplings of the $Z$ boson to quarks and $\theta$ is the center of mass scattering angle of the $W$ boson. Fig. \ref{WZtermX} shows the $d\bar u\to W^-Z$ squared helicity amplitude for $\lambda_W=-1$ and $\lambda_Z=+1$. 
\begin{figure}[t]
\begin{center}
\epsfysize=4 in
\epsffile{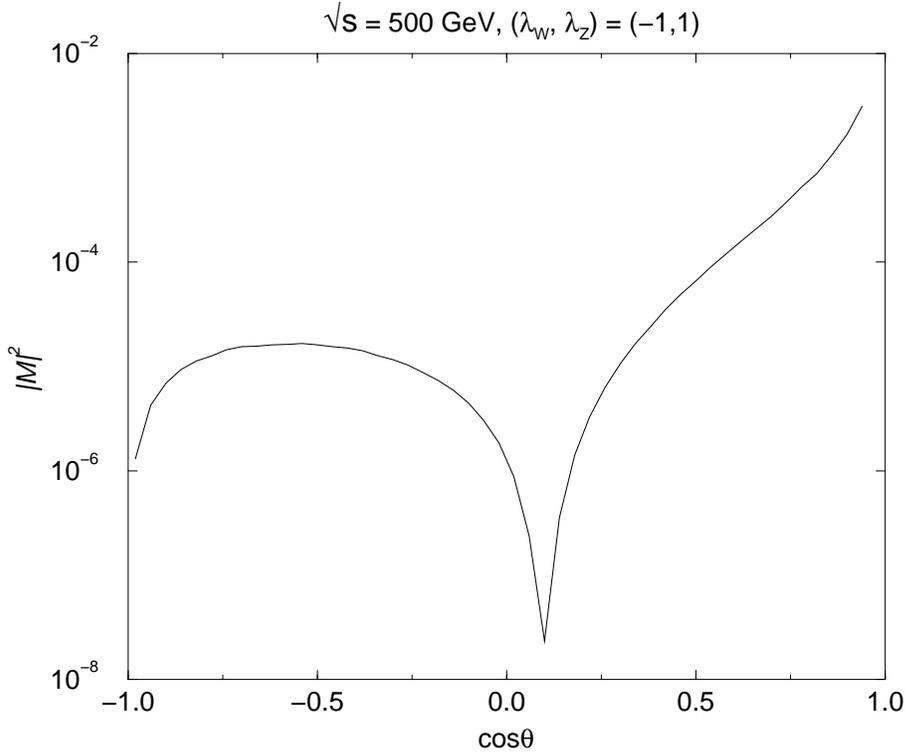} 
\end{center}
\caption{Amplitude zero in  $d \bar u \to W^- Z$ production.}
\label{WZtermX}
\end{figure}
The processes 
\begin{equation} 
e^{-} \bar \nu_{e} \to W^{-}Z~,~\nu_{e}e^{+} \to W^{+}Z
\end{equation}
also have similar zeros, as expected. 

There are additional $WZ$ production amplitude zeros, which are spin and energy dependent. The energy dependence is especially strong around the $WZ$ threshold. Most of these zeros are located in the  physical 
range of variables.  We located the positions of the zeros both using
$MATLAB$, algebraically, using explicit expressions for the amplitudes, and from the plots, $|M|^{2}~ vs~ \cos\theta~$ 
 (for the values of the  $\cos\theta$ in the physical region, 
$|\cos\theta| \leq 1$). In Fig. \ref{WZnew} we show several zeros of this 
type.
\begin{figure}
\begin{center}
\epsfysize=4in
\epsffile{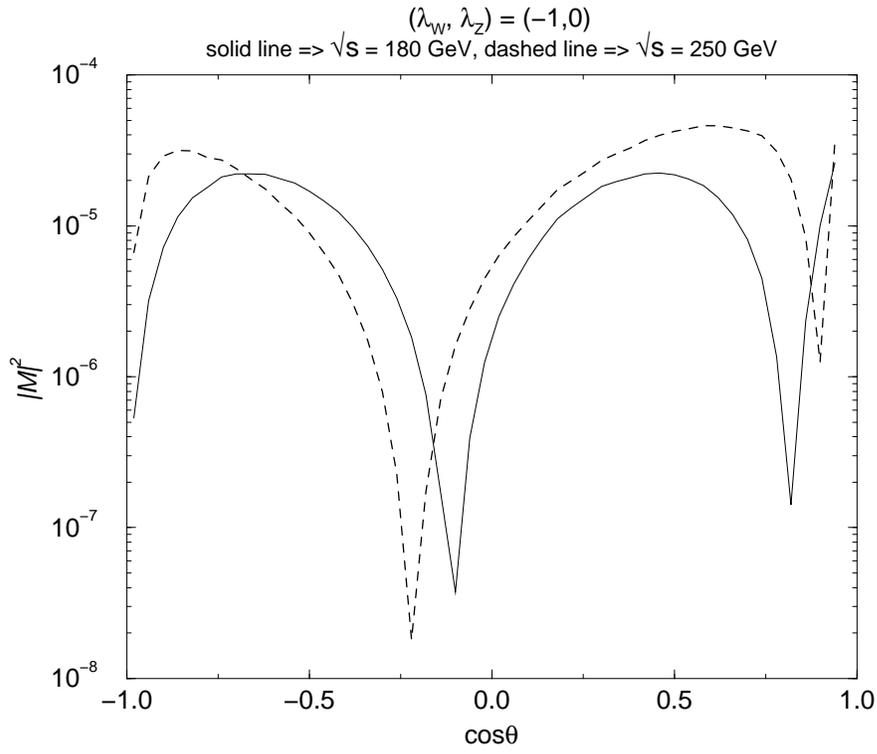} 
\epsfysize=4 in
\epsffile{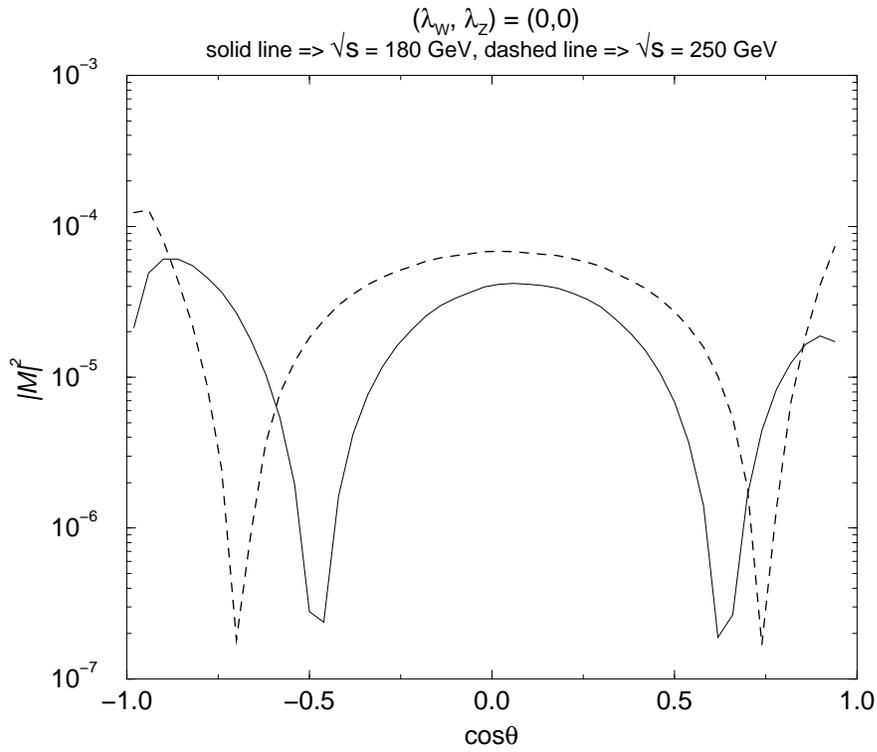} 
\end{center}
\caption{Zeros in  $d \bar u \to W^-Z$ production for specific helicity amplitudes.}
\label{WZnew}
\end{figure}
 Some of these zeros are approximately symmetric in $\cos \theta$ for  the most part of the c.m. energy range  (see Fig. \ref{WZsymm}) and leave  two symmetric dips in the rapidity distributions for $PP \to W^{-}Z$ . This occurs because two types of parton level contributions ($d~ quark$ (beam 1)+$\bar u~ quark$ (beam 2) and $d~ quark$ (beam 2)+$\bar u~ quark$ (beam 1) act coherently in terms of these zeros. Here `beam 1' and `beam 2' written in the parentheses show from which of two LHC proton beams the particular quark originates.
\begin{figure}[t]
\begin{center}
\epsfysize=4 in
\epsffile{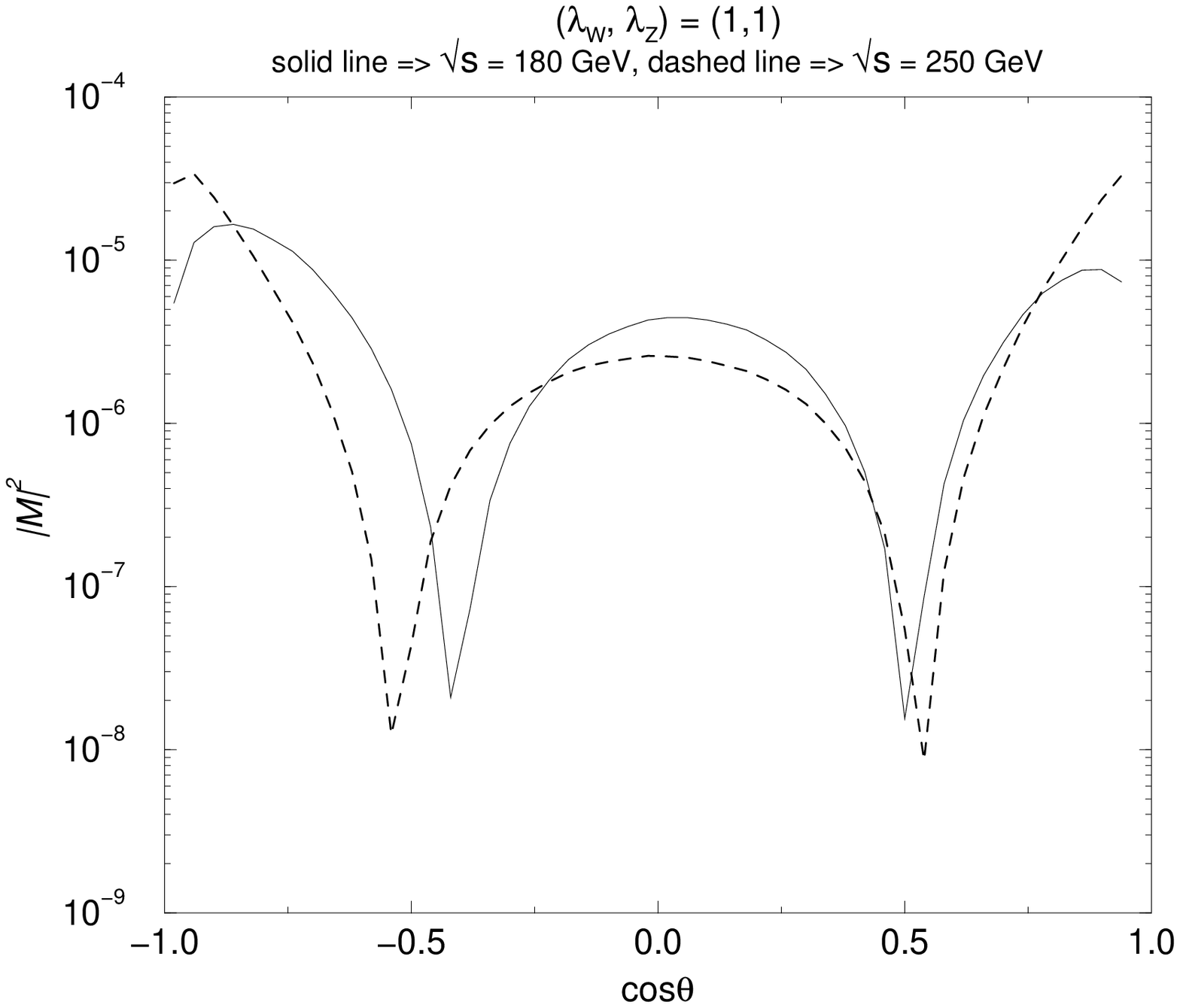} 
\end{center}
\caption{The squared $(\lambda_{W},\lambda_{Z})=(1,1)$
amplitude as a function of the scattering angle in the $d \bar u \to W^-Z$ 
production .}
\label{WZsymm}
\end{figure}
\begin{figure}
\begin{center}
\epsfysize=3.9in
\epsffile{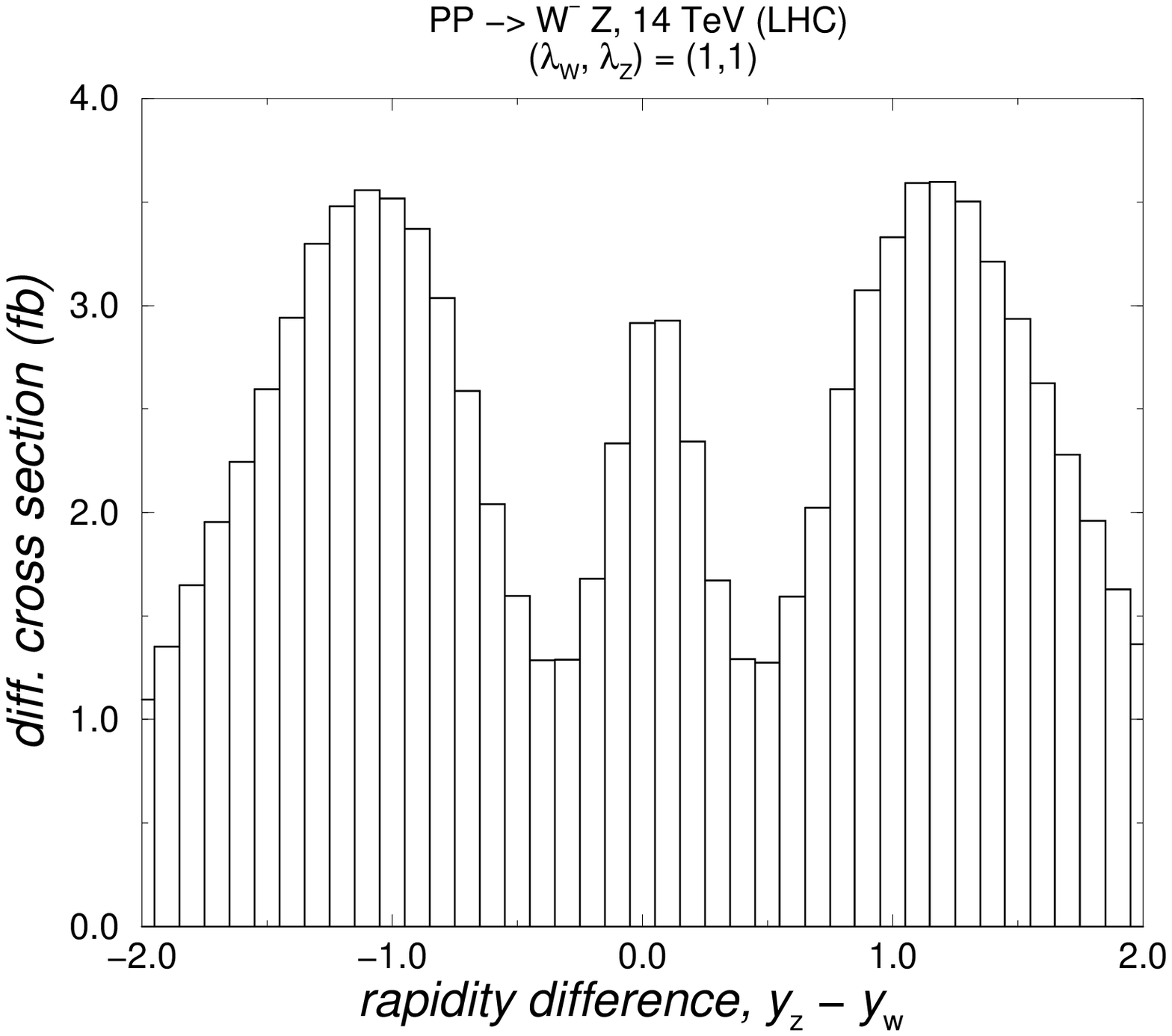} 
\epsfysize=3.9in
\epsffile{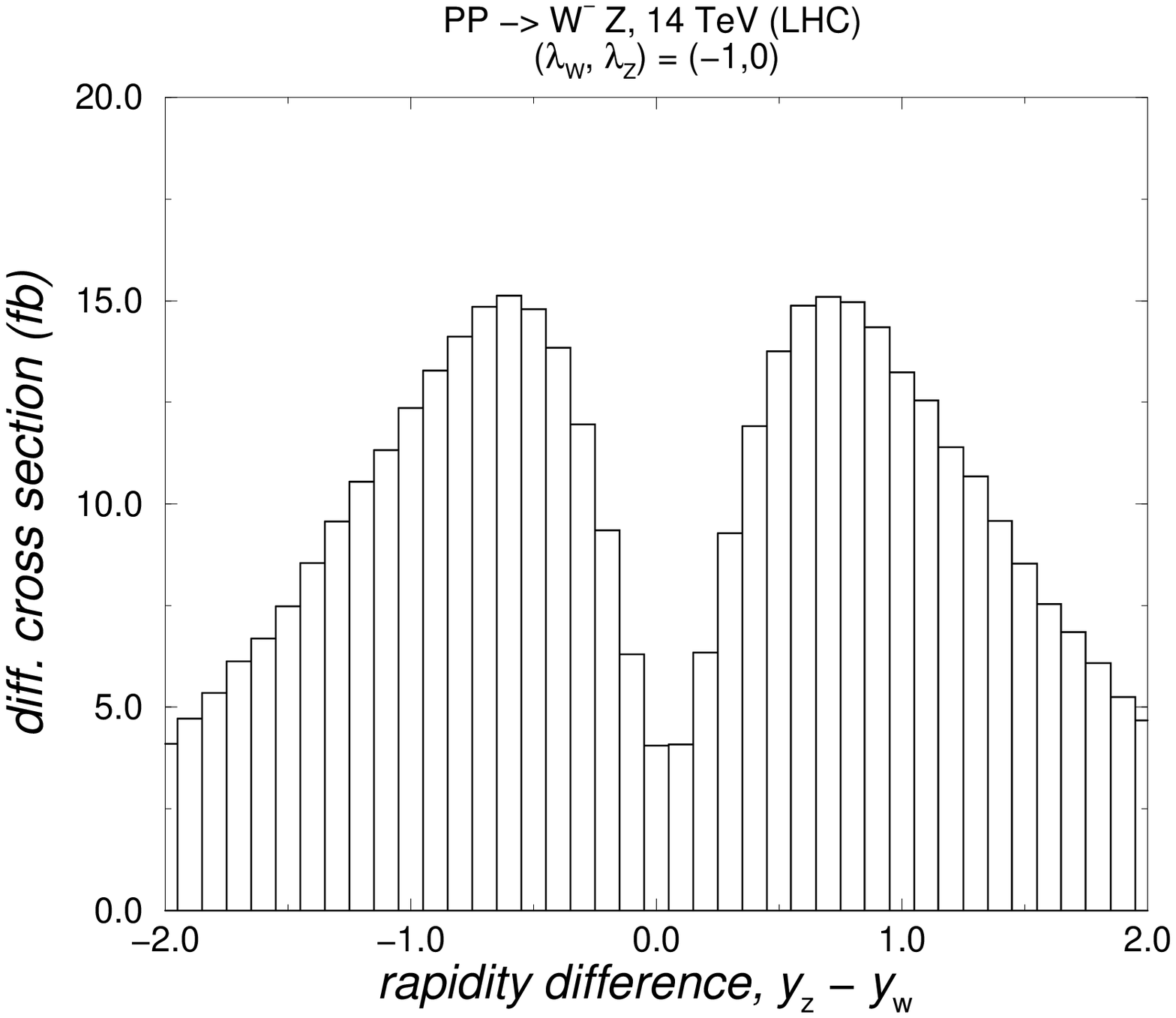} 
\end{center}
\caption{The $y_Z-y_W$ distribution for the $(\lambda_W,\lambda_Z)=
(1,1)$ and $(\lambda_W,\lambda_Z)=(-1,0)$  helicity amplitudes in $P P \to W^- Z$ at the LHC.}
\label{WZsymmP}
\end{figure}
These   zeros also leave deep dips in the rapidity distributions in the case of other helicity amplitudes, provided that they are relatively weakly energy dependent, so that the region of the dip in the $|M|^2$ vs  $\cos \theta$ distribution  due to the amplitude zero contains the value of $\cos \theta=0$ for all important values of the c.m. energy. The $(\lambda_W=-1,\lambda_Z=0)$ amplitude (Fig. \ref{WZnew}) is an example for this case\footnote{More than 90\% of the cross section for the $PP \to W^-Z$ with $~(\lambda_W=-1,\lambda_Z=0)~$ at 14 TeV originates from the parton level process $d \bar u \to W^- Z,~~(\lambda_W=-1,\lambda_Z=0)$ with the  c.m. energy $\sqrt{s} \le$ 350 GeV.}. 

The rapidity difference  $y_Z-y_W$ distributions at the LHC for these two helicity combinations of particles are shown in Fig. \ref{WZsymmP}.
\end{subsection}
\begin{subsection}{Zeros in $WZ\gamma$, $WZZ$  and  $WH\gamma$ 
production}
In  $WZ\gamma$, $WZZ$  and  $WH\gamma$ production all the nonzero charges have the same sign (see Eq. (\ref{Eq:Qsign})) and all three processes include at least one neutral gauge boson. Therefore one expects the existence of charge null zones for these processes. As we shall demonstrate in the following, all these processes exhibit zeros in the physical range of variables. 

The zeros of the  $d \bar u \to W^- Z\gamma$ production process amplitudes
are especially interesting, as the same helicity amplitude
may have zeros originating from different sources.

Since it is not possible to write down a simple analytical expression for the squared matrix element in the general case (see  Fig. \ref{WZgammaFeynman} for the $WZ\gamma$ production Feynman diagrams), we try to identify the zeros near the threshold values of the center of mass energy of the $WZ\gamma$ system, 
\begin{figure}
\begin{center}
\epsfysize=6in
\epsffile{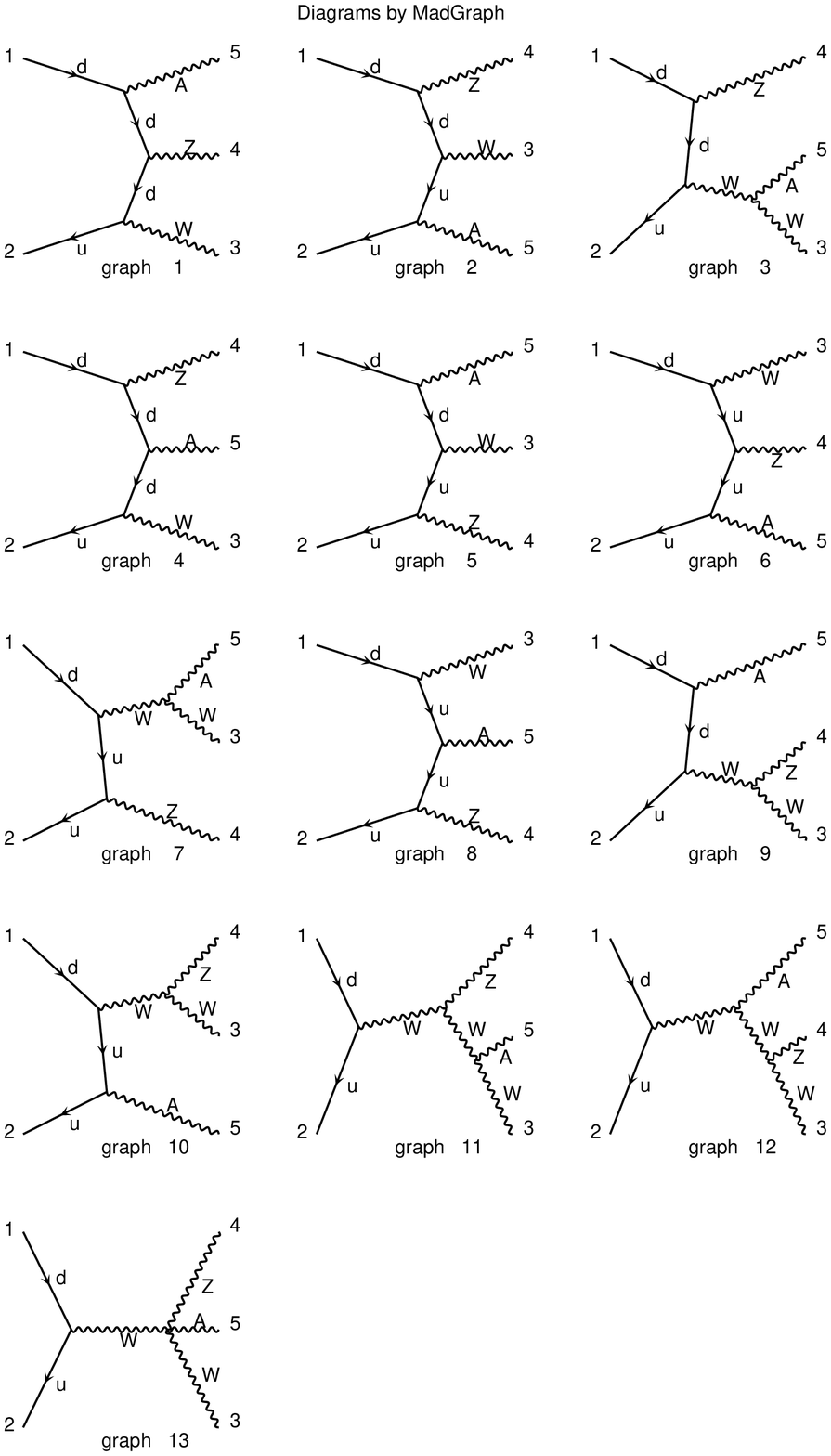}
\end{center}
\caption{ $WZ\gamma$ production Feynman diagrams generated by MadGraph.}
\label{WZgammaFeynman}
\end{figure}
\begin{equation}
\sqrt{s}~\approx~ M_{W}+M_{Z}~,
\end{equation}
where this can readily be done~\cite{Stirling_1}. In this limit the photon momentum is small compared to the initial fermion (quark) momenta and therefore can be neglected in the numerators of the internal fermion propagators.
It will be clear from the formulas below that if we choose the gauge, where transversely polarized (physical) photons have no 4th component, we can also neglect the contribution of the $WZ\gamma$ vertex to the amplitude, as both W boson and photon 3-momenta are very small. Under these conditions the  $W^{-}Z \gamma$ production amplitude can be written as    
\begin{equation}
M(W^{-}Z\gamma)=(-e)M(W^{-}Z)\epsilon^{*}_{\mu}(k_{\gamma})j^{\mu}~,
\end{equation}
where $j^{\mu}$ is given by
\begin{equation}
\label{Eq:Jmu}
 j^{\mu}=Q_{d}\frac{p_{d}^{\mu}}{p_{d}\cdot k_{\gamma}}+(-1-Q_{d})\frac{p_{\bar{u}}^{\mu}}{p_{\bar{u}}\cdot k_{\gamma}}- \frac{k_{W}^{\mu}}{k_{W}\cdot k_{\gamma}}~. 
\end{equation}
Here we considered the incoming quarks massless and used the Dirac equation to simplify the expression for  $j^{\mu}$.
Since the 4th component of the W boson momentum ($\approx{M_{Z}}$) is not small,  we keep $k_{W}^{\mu}$ in Eq. (\ref{Eq:Jmu}).

The condition 
\begin{equation}
\epsilon^{*}_{\mu}(k_{\gamma})j^{\mu}=0
\end{equation}
gives  the zero of the amplitude at $\cos\theta_{\gamma}=\frac{1}{3}$, where $\theta_{\gamma}$ is the angle between the incoming d quark and the photon. To see this, one can choose 
\begin{equation}
p_{d}=(E,0,0,E)~,~~
p_{\bar{u}}=(E,0,0,-E)~,~~
k_{\gamma}=E_{\gamma}(1,\sin\theta_{\gamma},0,\cos\theta_{\gamma})~,
\end{equation}
\begin{equation}
\epsilon^{\mu}_{1}=(0,0,1,0)~,~~~
\epsilon^{\mu}_{2}=(0,-\cos\theta_{\gamma},0,\sin\theta_{\gamma})~.
\end{equation}
 Then  
\begin{equation}
 \epsilon^{*}_{\mu}(k)j^{\mu}=-Q_{d}\frac{\sin\theta_{\gamma}}{1-\cos\theta_{\gamma}}+(-1-Q_{d})\frac{\sin\theta_{\gamma}}{1+\cos\theta_{\gamma}}=0
\end{equation}
or
\begin{displaymath}
 \cos\theta_{\gamma}=1+2Q_{d}= \frac{1}{3}~.
\end{displaymath}

 In the case when the Z boson and photon are collinear, 
\begin{equation}
\cos\theta_{W}=\cos(\theta_{\gamma}+\pi)=- \frac{1}{3}~.
\end{equation}
 
It is interesting to notice that the  $\cos\theta_{\gamma}$ distribution will have a zero in this limit, even if the $Z$ and $\gamma$ are not collinear\footnote{A similar situation occurs also in $~~~~d \bar{u} \to W^{-}\gamma\gamma~~~$, even for a larger range of center of mass energies, due to the masslessness of the photon.}, whereas  the $\cos\theta_{W}$ distribution exhibits the zero only if the $Z$ and $\gamma$ are collinear. Now it is clear that any other zero(s) should come from the first term, i.e. from the $M(W^{-}Z)$ amplitude.
Let's take, for example, the amplitude for $~(\lambda_W, \lambda_Z, \lambda_\gamma) = (0,-1,-1 )~$. The squared amplitude vs $\cos\theta_{W}$ for
the near the threshold value of the c.m. energy is given in Fig. \ref{wz&wza1}. 
We also show the $WZ$ production amplitude for $(\lambda_W, \lambda_Z) =(0,-1)$  in this figure. From the figure we see that, indeed,   the $WZ$ amplitude exhibits zeros at exactly the same locations as  the $WZ\gamma$ amplitude. 
\begin{figure}[t]
\epsfysize=4 in
\centerline{\epsffile{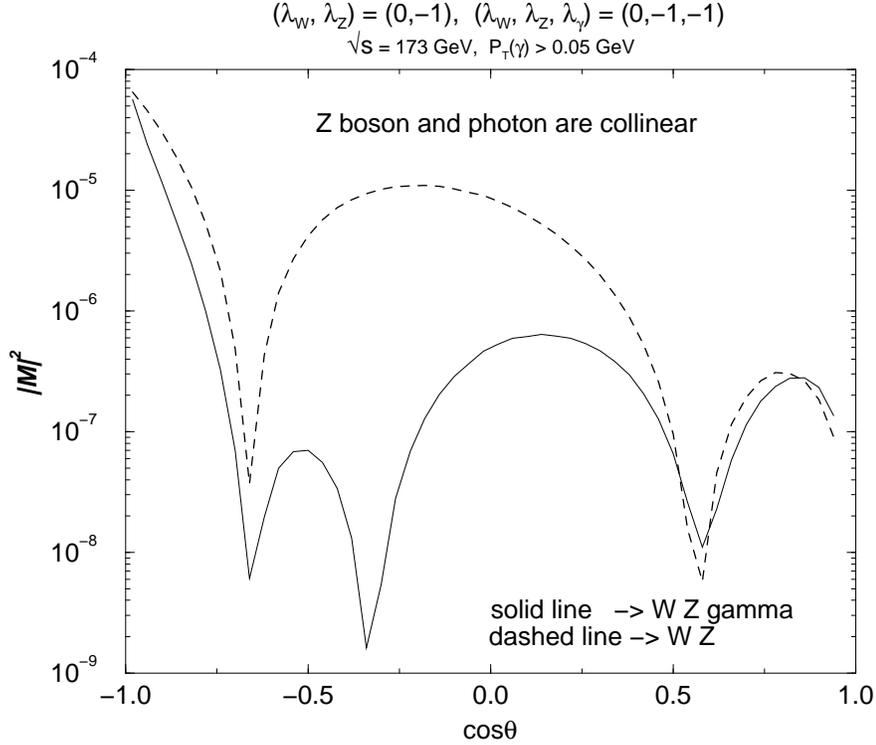}}
\caption{Zeros in $d\bar u\to W^-Z$ and  $d\bar u\to W^-Z \gamma$ for
a center of mass energy of $\sqrt{s}=173$ GeV. Shown are the squared 
amplitudes for $(\lambda_{W},\lambda_{Z})=(0,-1)$ and
$(\lambda_{W},\lambda_{Z},\lambda_{\gamma})=(0,-1,-1)$. A 
$p_{T}(\gamma)>0.05$ GeV cut has been imposed to avoid the infrared 
singularity present in $WZ\gamma$ production. The $Z$ boson and the
photon are collinear in the $WZ\gamma$ production case.}
\label{wz&wza1}
\end{figure}
By requiring that the $Z$ and $\gamma$ are collinear, it can be seen that 
that `-$\frac{1}{3}$' zeros (dips) are energy ($\sqrt{s}$) and spin independent. The positions of the zeros in the $WZ$ amplitudes and the corresponding zeros in the  $WZ\gamma$ amplitudes also remain the same at the increasing values of $\sqrt{s}$ for the collinear $Z$ and $\gamma$. 

 In the case when the $Z$ and $\gamma$ are collinear, the total momentum of the $Z \gamma$ system is given by
\begin{equation}
\label{Eq:ZG1}
\vec p=\vec p_{Z}+\vec p_{\gamma}=(1+\alpha) \vec p_{Z}~,
\end{equation}
where $\alpha=\frac{p_{\gamma}}{p_{Z}}$.
On the other hand, the invariant mass of the system is
\begin{equation}
\label{Eq:ZG2}
     M_{inv}=(E_{Z}+E_{\gamma})^2-(\vec p_{Z}+\vec p_{\gamma})^2=M_{Z}^2+2\alpha p_{Z} \sqrt{M_{Z}^2+p_{Z}^2}-\alpha p_{Z}^2~.
\end{equation}
By solving Eq. (\ref{Eq:ZG1}) and Eq. (\ref{Eq:ZG2}) together, we find 
\begin{equation}
\label{Eq:ZG3}
\alpha=\frac{-(Y^2+Yp^2) \pm \sqrt{(Y^2+Yp^2)^2-Y^2(Y^2-M_{Z}^2p^2)}}{Y^2-M_{Z}^2p^2}~,
\end{equation}
where 
\begin{equation}
Y=\frac{M_{inv}^{2}-M_{Z}^{2}}{2}~.
\end{equation}  
However, for this value of $\alpha$, the momenta could be either collinear or opposite in direction, depending on whether the sign of $\alpha$ is positive or negative (see Eq.(\ref{Eq:ZG1})).

We can remove the latter combination of $p_{Z}$ and $p_{\gamma}$ by requiring that
\begin{equation}
\label{Eq:ZG4}
Y^2-M_{Z}^2p^2 \leq 0
\end{equation}
and it is clear that only the ``-'' solution of Eq. (\ref{Eq:ZG3}) should be used for the collinear case.

 Eq. (\ref{Eq:ZG4}) can be transformed into a 4th order equation in   $M_{inv}$ and used to determine the upper limit of  the invariant mass $M_{inv}$, 
when the $Z$ boson and the photon are still collinear:
\begin{equation}
\label{Eq:Minv}
\Big ( 1-\frac{M_{Z}^2}{s} \Big) M_{inv}^4+2\frac{M_{Z}^2M_{W}^2}{s}M_{inv}^2-M_{Z}^2 {s}-\frac{M_{Z}^2M_{W}^4}{s}+2M_{Z}^2M_{W}^2+M_{Z}^4 \leq 0~.
\end{equation}
Here $s$ is the squared center of mass energy of the $WZ\gamma$ system.

The solutions of Eq. (\ref{Eq:Minv}) can readily be found by $MATLAB$. In table \ref{tableMinvWZg} we give the values of these upper limits for the invariant mass of the collinear $Z \gamma$ system, $M_{inv}^{max}$, at the different values of c.m. energy $\sqrt{s}$.
\begin{table}
\begin{center}
\begin{tabular}{|c|c|c|c|c|c|c|} \hline
$\sqrt { s}$ & \ 173 & \ 200 & \ 250 & \ 300 & \ 350 & \ 400 \\ \hline
$M_{inv}^{max}$ & \ 92.63 & \ 113.25 & \  138.16 & \ 156.64 & \  172.16 & \ 185.92  \\ \hline  
\end{tabular}
\end{center}
\caption{ {\small The maximum invariant mass  $M_{inv}^{max}$ for 
the collinear $Z$ boson and $\gamma$ system
at the different c.m. energies $\sqrt{s}$ for $WZ\gamma$ production. All energy values are in units of GeV.}}
\label{tableMinvWZg}
\end{table}

The other helicity amplitudes of  $WZ\gamma$ production also
have zero-rich structures. The amplitudes ($\pm 1,\mp 1$,-1), ($\pm 1,\mp 1$,+1) 
exhibit the earlier mentioned zeros of $WZ$ production (see Eq. (\ref{Eq:Xtermzero}) and the text before it) and `-$\frac{1}{3}$' 
zero. 

We would like to emphasize that the position of the `-$\frac{1}{3}$' dip  may vary slightly and the depth of the dip may change substantially, depending on the transverse momentum cut of the photon and c.m. energy, revealing its approximate nature.

To summarize, the $WZ\gamma$ production amplitudes have a very 
rich structure of zeros originating from three different sources:
\\
%\begin{enumerate}
1. a radiation zero at $\cos\theta_{\gamma}= \frac{1}{3}$ connected with
photon radiation,
\\
2. the approximate radiation zero occurring in $WZ$ production,
\\
3. spin dependent zeros in some of the $WZ$ production amplitudes 
discussed above.
\\
%\end{enumerate} 
The zero present at $\cos\theta_{\gamma}= \frac{1}{3}$ leads to a clear dip in
the $~y_{Z\gamma}-y_{W}~$ rapidity difference distribution ($y_{Z\gamma}$ is 
the rapidity of the $Z\gamma$ system), if the cosine of the angle between 
the $Z$ boson and photon is restricted to $\cos\theta(Z,\gamma)>0$. The 
$y_{Z\gamma}-y_{W}$ distribution at the LHC is shown in Fig. \ref{wzay}.
\begin{figure}[t]
\epsfysize=4 in
\centerline{\epsffile{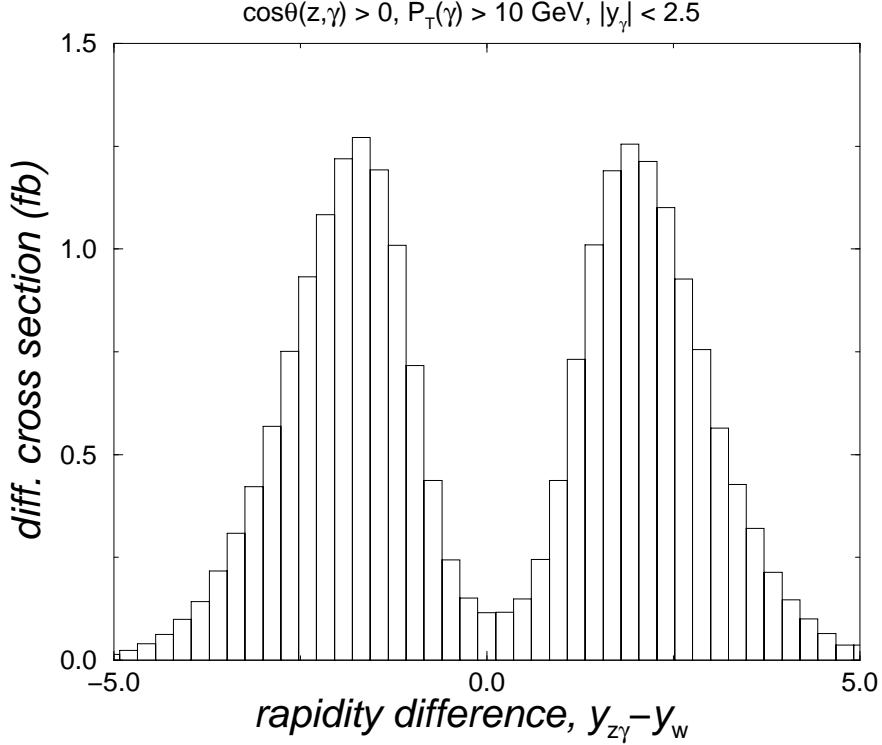}}
\caption{The $y_{Z\gamma}-y_{W}$ rapidity difference distribution for
$\cos\theta(Z,\gamma)>0$ in $WZ\gamma$ production at the LHC.}
\label{wzay}
\end{figure}

The $WH\gamma$ production amplitudes also have a `-$\frac{1}{3}$' zero, if 
the Higgs boson $H$ and $\gamma$ are collinear (Fig. \ref{wha1}).  The rapidity dip for  $WH\gamma$ production distribution is not as prominent as in the 
$WZ\gamma$ case due to the lack of the $WZ$ zeros present in the $WZ\gamma$
helicity amplitudes\footnote{$WH$ as well as $ZH$ amplitudes do not exhibit 
amplitude zeros of any of the types considered in this section, as expected.}.

\begin{figure}
\begin{center}
\epsfysize=4in
\epsffile{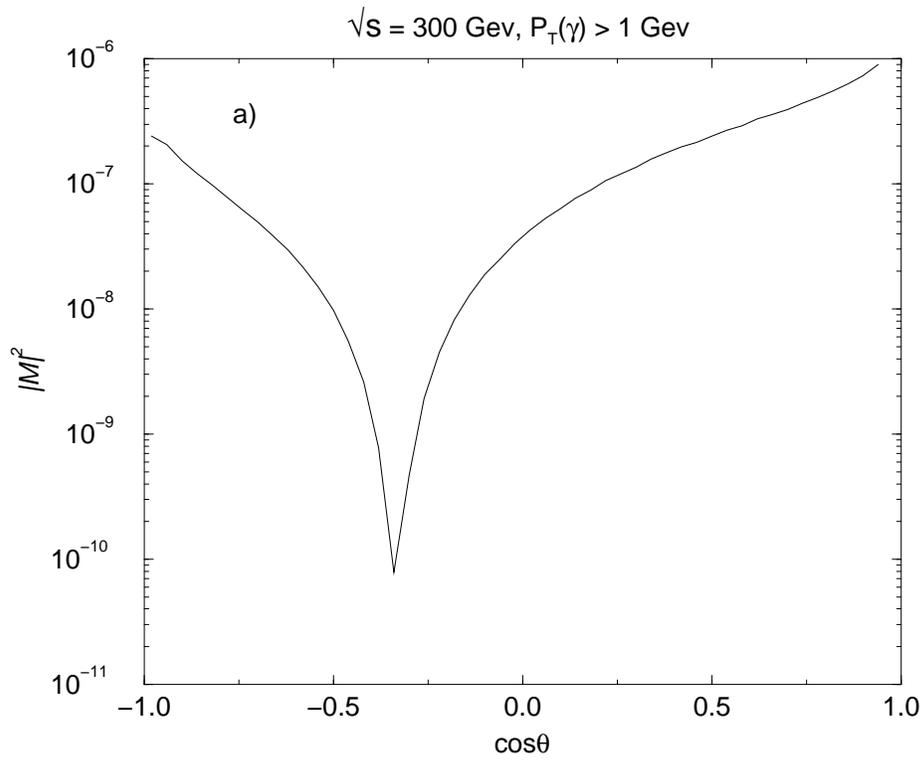} 
\epsfysize=4in
\epsffile{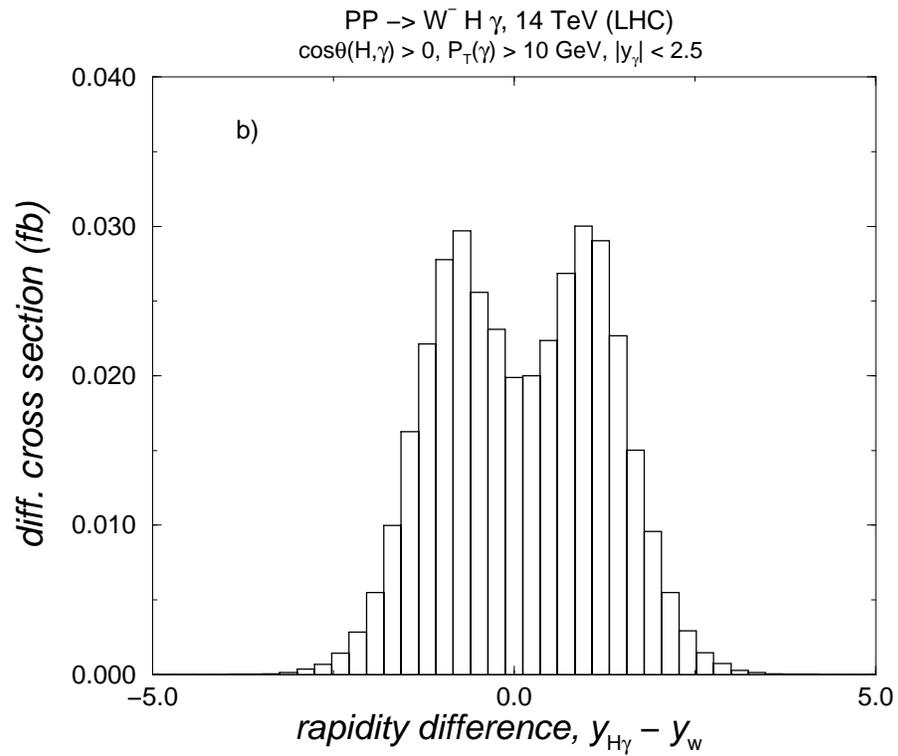}
\end{center}
\caption{ a) Zeros in $d \bar u \to W^- H \gamma$ production,  b) The dip in the rapidity  distribution in  $PP \to W^-H\gamma$.}
\label{wha1}
\end{figure}
Similar to $WZ$ production, 
the amplitudes for $WZZ$ production exhibit an approximate radiation
zero, if the two $Z$ bosons are collinear (Fig. \ref{wzz141}). 
\begin{figure}
\begin{center}
\epsfysize=4 in
\epsffile{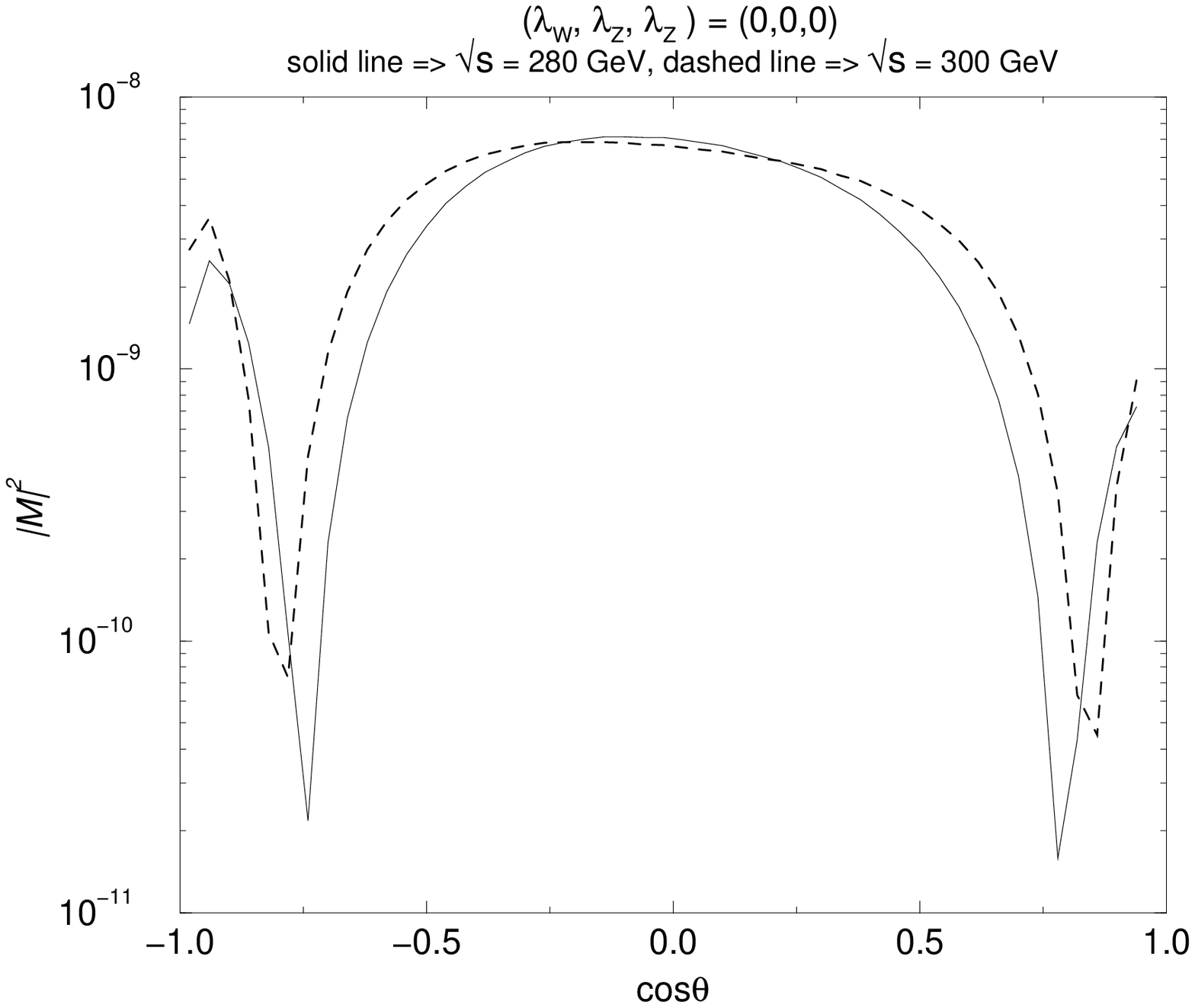} 
\end{center}
\caption{Zeros in $d \bar u \to W^-ZZ$ production. Shown is the amplitude 
with all three gauge bosons longitudinally polarized}
\label{wzz141}
\end{figure}
The technique for 
generating collinear $Z$ bosons is similar to that of the $Z\gamma$ ($H\gamma$) case with the only significant difference that both ``-'' and ``+'' solutions of the equation for $\alpha$ (see the text after Eq. (\ref{Eq:ZG4})) can generate the collinear $Z$ bosons. The amplitude zeros lead to a 
dip in the 
$y_{ZZ}-y_{W}$ distribution, which, however, is much less pronounced than
the corresponding dip in the $y_{Z\gamma}-y_{W}$ distribution in
$WZ\gamma$ production, due to the absence of the  `-$\frac{1}{3}$' zero in the $W ZZ$ amplitudes. A comparison of the two distributions at the LHC 
is shown in Fig. \ref{wza&wzz}. We only considered the contributions of the 
amplitudes, where the first two particles ($W$ and $Z$) have the same helicities as the dominating helicity amplitudes of the $WZ$ production, $(\lambda_W,\lambda_Z)=(\pm 1, \mp 1)$. There are four such amplitudes in the $WZ\gamma$ case and six amplitudes in the $WZZ$ case.  These amplitudes give a substantial part of the contributions to the total cross sections, while the rapidity difference distributions due to these amplitudes still exhibit clear dips without any restriction on the angle between the $Z$ boson and photon (the second $Z$ boson).
\begin{figure}
\begin{center}
\epsfysize=3.9 in
\epsffile{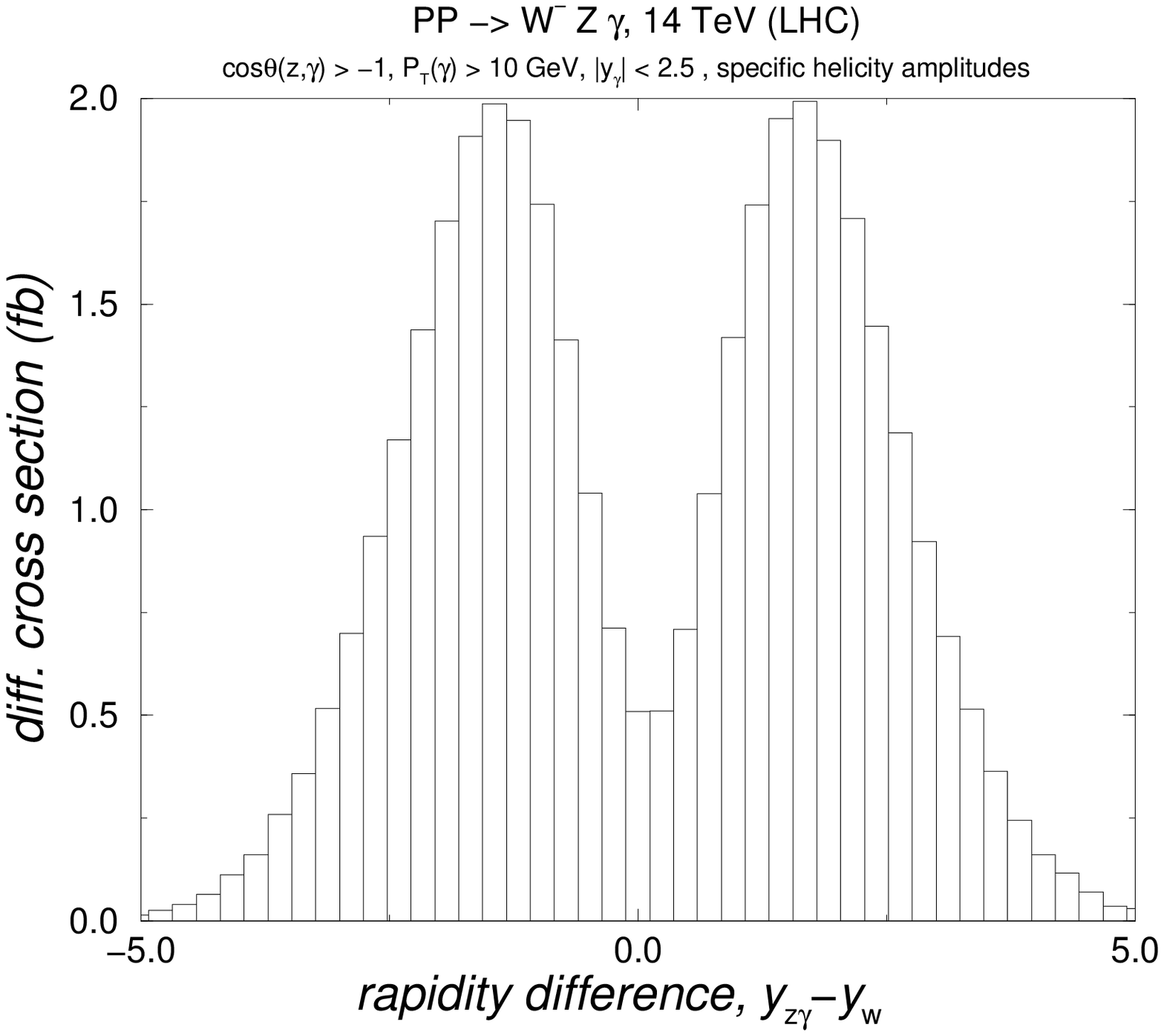} 
\epsfysize=3.9 in
\epsffile{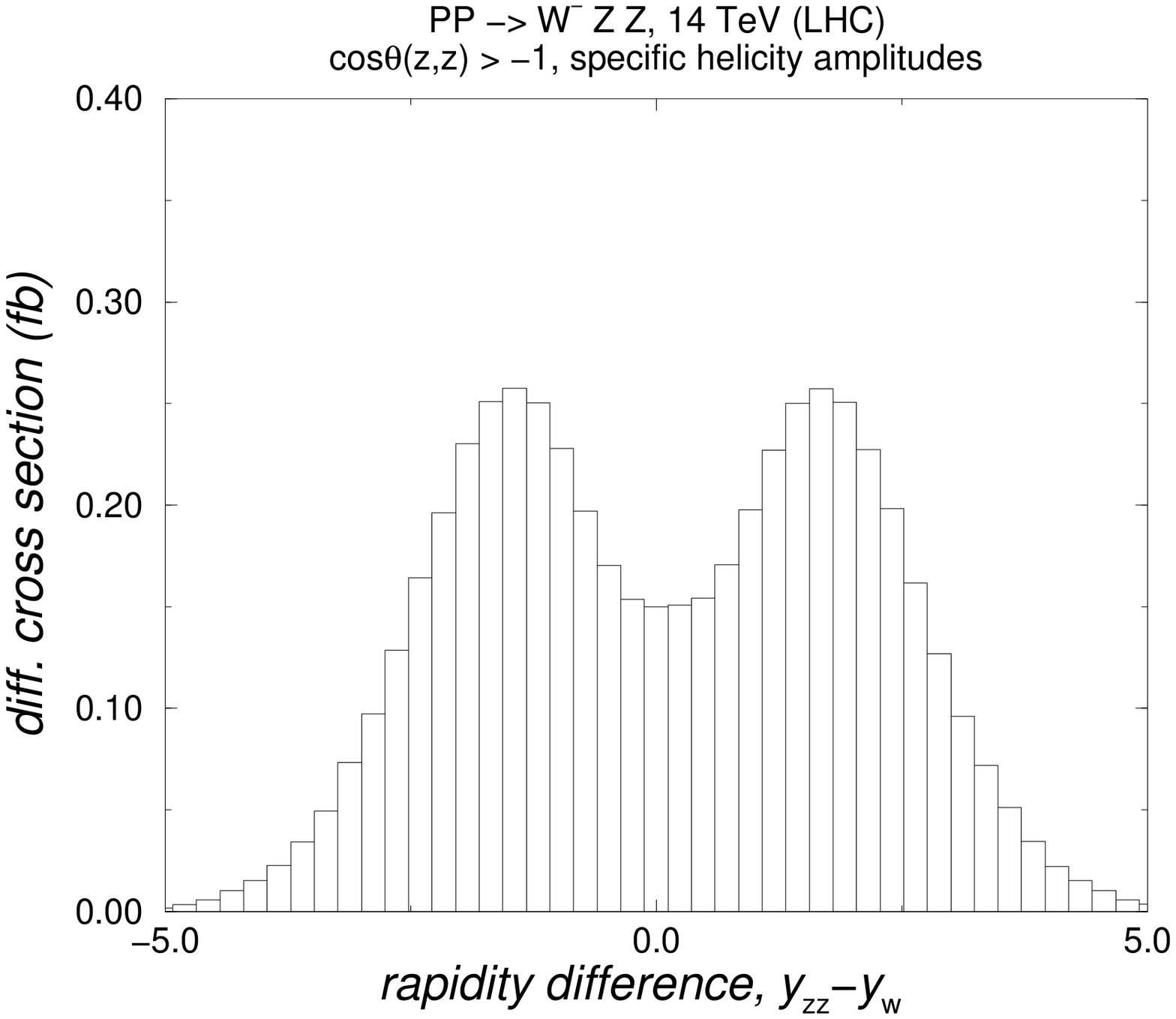}
\end{center}
\caption{Comparison of the $y_{Z\gamma}-y_{W}$ distribution in
$WZ\gamma$ production, and the $y_{ZZ}-y_{W}$ distribution in $WZZ$
production at the LHC. Contributions of the specific helicity amplitudes (see text) are considered.}
\label{wza&wzz}
\end{figure}

The amplitudes for none of these three  production processes exhibit the current
null zone zeros discussed in the $W\gamma$ case.
\end{subsection}

\begin{subsection}{Zeros in neutral electroweak boson production processes
and the helicity amplitude relations}
\indent Since these  processes require the initial
particles to have charges of different sign, the possibility of a charge 
null zone is excluded in this case (see Eq. (\ref{Eq:Qsign})). Nevertheless
some of these processes, namely, $Z \gamma$, $Z H \gamma$ and  $Z Z \gamma$ 
production exhibit the same current null zones as the $W \gamma$ production 
process.  
The amplitudes vanish \textit{for all scattering angles} within the SM, 
when the $Z$
 boson is longitudinally polarized and the photon polarization vector 
$~\vec \epsilon~$
is perpendicular to the scattering plane in the c.m. frame. In the $Z Z \gamma$ case, these are the helicity amplitudes for which both of the $Z$ bosons 
are longitudinally polarized. 

The $Z H \gamma$ production amplitudes exhibit zeros only if $H$ and $\gamma$ are collinear, while  the  $Z Z \gamma$ production amplitudes will still have approximate current null zone zeros for nonzero values of the angle 
between $H $ and $\gamma$.

The existence of the current null zones discussed above implies 
the identities for the helicity amplitudes. For example,  the current null zone in $Z\gamma$ production
is the result of the following equality of the helicity amplitudes,
\begin{equation}
\label{Eq:Mzgamma}
M(\lambda_Z=0,\lambda_\gamma=-1)=M(\lambda_Z=0,\lambda_\gamma=1)~.
\end{equation}
 The two polarization 
amplitudes built out of these helicity amplitudes are
\begin{equation}
M(\lambda=1)=\frac{i}{\sqrt{2}}(M(\lambda_Z=0,\lambda_\gamma=1)+M(\lambda_Z=0,\lambda_\gamma=-1))
\end{equation}
and 
\begin{equation}
M(\lambda=2)=\frac{1}{\sqrt{2}}(M(\lambda_Z=0,\lambda_\gamma=1)-M(\lambda_Z=0,\lambda_\gamma=-1))~.
\end{equation}
The polarization  amplitude $M(\lambda=2)$ will have a zero value for all 
scattering angles due to Eq. (\ref{Eq:Mzgamma}), which is the current null 
zone of the $Z\gamma$ production process.

Similarly, the current null zone in  $ZZ\gamma$ production is due to
the helicity amplitudes $M(\lambda_Z=0, \lambda_Z=0, \lambda_\gamma=\pm 1)$ being equal.

The equality of different helicity amplitudes does not always lead to the existence of a current null zone. In the $Z\gamma \gamma$ case,
one can show that
\begin{equation}
 M(\lambda_Z=0, \lambda_\gamma=-1, \lambda_\gamma=-1)=
 M(\lambda_Z=0, \lambda_\gamma=1, \lambda_\gamma=1)~
\end{equation}
and 
\begin{equation}
 M(\lambda_Z=0, \lambda_\gamma=-1, \lambda_\gamma=1)=
 M(\lambda_Z=0, \lambda_\gamma=1, \lambda_\gamma=-1)~.
\end{equation}
Similar to the $W\gamma \gamma$ case, since both photons have 
different helicities 
for these amplitudes, they cannot be combined into polarization
amplitudes, one of which would have a zero value. Therefore, surprisingly,
 $Z\gamma \gamma$ production does not  have a current null zone of 
the type found in the  $Z \gamma$ and $ZZ \gamma$ cases. 

The equality of the amplitudes with different particle helicities
is also observed for the $WW$ and $ZZ$ production amplitudes:
\begin{equation}
 M(\lambda_{W(Z)}=-1, \lambda_{W(Z)}=-1)=
 M(\lambda_{W(Z)}=1, \lambda_{W(Z)}=1)~,
\end{equation} 
\begin{equation}
 M(\lambda_{W(Z)}=0, \lambda_{W(Z)}=-1)=
 M(\lambda_{W(Z)}=1, \lambda_{W(Z)}=0)~.
\end{equation} 
Therefore, one can consider the current null 
zone, occurring at all scattering angles, as a special case of this
equality of helicity 
amplitudes with different particle helicities, it occurs when 
the helicity of only one of the particles has  different values 
for two amplitudes which are equal and it is therefore possible to combine
the two amplitudes into the polarization amplitudes, one of which has
 zero value.

It is interesting that such a `classical' process, as the two photon
 production process, for example, $u \bar u \to \gamma \gamma$, also 
exhibits  the equalities of the amplitudes,
and this occurs when all four particles have different helicities 
in two compared amplitudes:
\begin{displaymath}
 M(\lambda_{u}=-1, \lambda_{\bar u}=1,\lambda_{\gamma}(1)=-1~, 
\lambda_{\gamma}(2)=-1)= 
\end{displaymath}
\begin{equation}
M(\lambda_{u}=1, \lambda_{\bar u}=-1,\lambda_{\gamma}(1)=1, \lambda_{\gamma}(2)=1)~,
\end{equation} 
\begin{displaymath}
 M(\lambda_{u}=-1, \lambda_{\bar u}=1,\lambda_{\gamma}(1)=-1~, 
\lambda_{\gamma}(2)=1)= 
\end{displaymath}
\begin{equation}
M(\lambda_{u}=1, \lambda_{\bar u}=-1,\lambda_{\gamma}(1)=1, \lambda_{\gamma}(2)=-1)~,~ etc.
\end{equation}   

In table \ref{tablenullzones} we present a summary of TYPE I null zones that we discussed in this section. We only included the current null zones that occur for all values of the scattering angle.

\begin{table}
\begin{center}
\begin{tabular}{|c|c|c|c|c|} \hline
{\small process}&\ {\small Charge null zone} & \ {\small Current null zone} & \ {\small Helicity amplitudes with equal values} \\ \hline 
$W\gamma$ & \ + & \ + & \ + \\ \hline
$W\gamma\gamma$ & \ + & \ - & \ + \\ \hline
$WZ$ & \ + & \ - & \ - \\ \hline
$W Z \gamma$ & \ + & \ - & \ - \\ \hline
$W H \gamma$ & \ + & \ - & \ - \\ \hline
$W Z Z$ & \ + & \ - & \ - \\ \hline
$Z\gamma$ & \ - & \ + & \ + \\ \hline
$Z H \gamma$ & \ - & \ + & \ + \\ \hline
$Z \gamma \gamma$ & \ - & \ - & \ + \\ \hline
$Z Z \gamma$ & \ - & \ + & \ + \\ \hline
$WW$ & \ - & \ - & \ + \\ \hline
$ZZ$ & \ - & \ -  & \ + \\ \hline
$\gamma \gamma$ & \ - & \ - & \ + \\ \hline
\end{tabular}
\end{center}
\caption{ {\small \textit{Charge null zone} zeros, \textit{current null zone} zeros and equalities of the helicity amplitudes exhibited by electroweak sector high energy processes within the SM in the physical region of parameters. `+' (`-') sign corresponds to the presence (absence) of the null zone or of the helicity amplitudes with the equal values}}
\label{tablenullzones}
\end{table}
\end{subsection} 
\section{CONCLUSION}
The SM amplitudes for processes  with the emission of one or more neutral gauge bosons exhibit zeros, the amplitudes vanish under the specific conditions. There are   many different forms of zeros.  The most investigated form of the zeros are TYPE I zeros. There are two forms of TYPE I zeros, the charge null zones and the  current null zones. In the case of TYPE I charge null zones, the distributions of the scattering angles contain zeros. In the case of the TYPE I current null zones, either the amplitude completely vanishes or the distributions for the scattering angles contain zeros for certain helicity combinations of the particles. We briefly discussed the origin of TYPE I zeros, which are due to the factorization of the production amplitudes. The amplitudes can also vanish for many processes, when the photon (gauge boson ) momentum  is in the scattering plane created by the momenta of the other particles, which participate in the process. These zeros are called TYPE II zeros.  Finally, there are  those amplitude zeros, such as the recently observed zeros in $WZ$ production, which could be just accidental zeros, due to the cancellations between the various terms in the amplitude. We have discussed all these different types of zeros and shown that some of the production amplitudes have an especially rich structure in terms of zeros. $WZ\gamma$ and $WZZ$ production amplitudes are examples for this. Many of the zeros leave deep dips in the rapidity distributions, which could be observed experimentally. We also showed that the TYPE I current zones, occurring for all scattering amplitudes, are the special case of the equality of the values of the production (helicity) amplitudes for the specific helicity combinations of the particles. 
\end{section}
\section*{ACKNOWLEDGMENTS}
The discussions of the subject of this paper with U. Baur and A. Werthenbach were helpful. The author appreciates the discussion  of the value of the 
vector-boson magnetic moment with M. Veltman. The author would also like to thank R. Gonsalves for his valuable comments during preparation of this work for publication.


\begin{references}
\bibitem{W_moment_1} T.D. Lee, C.N. Yang, \textit{Phys. Rev.}  \textbf{128},
885-898 (1962).
\bibitem{W_moment_2} K.J. Kim, Yung-Su Tsai,  \textit{Phys. Rev.}  \textbf{D7}, 3710-3721 (1973).  
\bibitem{Rad_0} K.O. Mikaelian, M.A. Samuel, and D. Sahdev, 
\textit{Phys. Rev. Lett.} \textbf{43}, 746-749 (1979).
\bibitem{Baur_1} U. Baur, S. Errede, and G. Landsberg, \textit{Phys. Rev.}\ \textbf{D50}, 1917-1930 (1994). 
\bibitem{Brown_1} R.W. Brown, K.L. Kowalski, and S.J. Brodsky,
\textit{Phys. Rev.} \textbf{D28}, 624-649 (1983).
\bibitem{Zhu} D. Zhu, \textit{Phys. Rev. } \textbf{D22}, 2266-2274 (1980).
\bibitem{Goebel} C.J. Goebel, F. Halzen, and J.P. Leveille, 
\textit{Phys. Rev.} \textbf{D23}, 2682-2685 (1980).
\bibitem{Brown_2} S.J. Brodsky,  and R.W. Brown,
\textit{Phys. Rev. Lett.} \textbf{49}, 966-970 (1982).
\bibitem{Brown_4} R.W. Brown, and K.L. Kowalski, \textit{Phys. Rev. } 
\textbf{D30}, 2602-2607 (1984).
\bibitem{Brown_5} R.W. Brown, and K.L. Kowalski, \textit{Phys. Rev. Lett.} 
\textbf{51}, 2355-2358 (1983).
\bibitem{Robinet_1} V. Barger, R.W. Robinett, W.Y. Keung, R.J.N. Phillips,
 \textit{Phys. Lett.} \textbf{B131}, 372-376 (1983). 
\bibitem{Brown_6}  R.W. Brown, and K.L. Kowalski, \textit{Phys. Lett.} 
\textbf{B144}, 235-239 (1984).
\bibitem{DELANEY_1} D. DeLaney, E. Gates, O. Tornkvist,  \textit{Physics Letters} \textbf{B186}, 91-95 (1987).
\bibitem{Brown_7} R.W. Brown, \textit{Vector Boson Symp. 1995}, 261-272. 
\bibitem{Brown_8} R.W. Brown, M.E. Convery, and M. A. Samuel, 
\textit{Phys. Rev. } \textbf{D49}, 2290-2297 (1994).
\bibitem{Heyssler_1} M. Heyssler, and W.J. Stirling, \textit{Eur. Phys. J.} 
\textbf{C4}, 289-299 (1998).   
\bibitem{Heyssler_2} M. Heyssler, and W.J. Stirling, \textit{Eur. Phys. J.} 
\textbf{C5}, 475-484 (1998). 
\bibitem{Stirling_1} W.J. Stirling, A. Werthenbach, 
\textit{ Eur. Phys. J. } \textbf{C12}, 441-450 (2000). 
\bibitem{Stelzer_1} T. Stelzer, W.F. Long,  \textit{Comput. Phys. Commun.} \textbf{81}, 357-371 (1994).
\bibitem{Cortes_1} J. Cortes, K. Hagiwara, F. Herzog, \textit{Phys. Rev. } 
\textbf{D28}, 2311-2313 (1983).
\bibitem{Brown_3} R.W. Brown, and K.L. Kowalski, \textit{Phys. Rev. } \textbf{D29}, 2100-2104 (1984).
\bibitem{Baur_2} U. Baur, T. Han, N. Kauer, R. Sobey, and D. Zeppenfeld, 
~\textit{Phys. Rev. } \textbf{D56}, 140-150 (1997). 
\bibitem{Han_Sobey} T. Han, R. Sobey, \textit{Phys. Rev.} \textbf{D52}, 6302-6308 (1995). 
\bibitem{Baur_3} U. Baur, T. Han, and J. Ohnemus, \textit{Phys. Rev. Lett.}
\textbf{72}, 3941-3944 (1994). 
\bibitem{Han} T. Han,  \textit{Vector Boson Symp. 1995}, 224-238.  
\end{references}
\end{document}